\documentclass[%
 reprint,
superscriptaddress,
 aps,pre,
]{revtex4-1}

\usepackage{color}
\usepackage{dcolumn}
\usepackage{bm}

\usepackage{amsmath}
\usepackage{amssymb}
\usepackage{graphicx}
\newcommand{\vect}[1]{\mathbf{#1}}

\newcommand{\vecr}{\vect{r}}

\newcommand{\rme}{\mathrm{e}}

\newcommand{\Vext}{V_\mathrm{fb}}
\newcommand{\Vextt}{\tilde{V}_\mathrm{fb}}

\newcommand{\VYu}{\Phi}

\newcommand{\rhop}{\rho_\mathrm{p}}

\newcommand{\tp}{t_\tau}
\usepackage{placeins}
\usepackage{dsfont}
\newcommand{\dif}{\mathrm{d}}

\begin{document}
\raggedbottom
\preprint{APS/123-QED}

\title{Traveling band formation in feedback-driven colloids}

\author{Sonja Tarama}
\email{sonja.tarama@hhu.de}
\affiliation{Institute for Theoretical Physics II: Soft Matter, Heinrich Heine University D\"usseldorf, Universit\"atsstra\ss e 1, D-40225 D\"usseldorf, Germany.}
 
\author{Stefan U. Egelhaaf}
\affiliation{Condensed Matter Physics Laboratory, Heinrich Heine University D\"usseldorf, Universit\"atsstra\ss e 1, D-40225 D\"usseldorf, Germany}
 
\author{Hartmut L\"owen}%
\affiliation{Institute for Theoretical Physics II: Soft Matter, Heinrich Heine University D\"usseldorf, Universit\"atsstra\ss e 1, D-40225 D\"usseldorf, Germany.}

\date{\today}

\begin{abstract}
Using simulation and theory we study the dynamics of a colloidal suspension in two dimensions subject to a time-delayed repulsive feedback 
that depends on the positions of the colloidal particles. The colloidal particles experience an additional potential that is a superposition of repulsive potential energies centered around the positions of all the particles a delay time ago. Here we show that such a feedback leads to
self-organization of the particles into traveling bands. The width of the bands and their
propagation speed can be tuned by the delay time and the range of the imposed repulsive potential.
The emerging traveling band behavior is observed in Brownian dynamics computer simulations as well as microscopic dynamic density functional theory (DDFT). Traveling band formation also persists in systems of finite size leading to rotating traveling waves in the case of circularly confined systems.
\end{abstract}

\maketitle


\section{Introduction}
Nonequilibrium systems subject to a feedback potential have been studied extensively in recent times 
\cite{Granger2016,Loos2014,Loos2017,Loos2018,Lospichl2018,Lopez2008,Gernert2015,Popli2018, Yang2018,Lichtner2010}. Due to the feedback, used e.g.~to stabilize 
dynamics \cite{Lichtner2010,Pyragas1992,Hoevel2005, Cohen2005,Jun2012} or structure \cite{Popli2018,Khadka2018}, the system dynamics becomes 
history-dependent. The feedback can be realized through external programming of a laser trap 
\cite{Popli2018, Blickle2011,Hanes2009,Evers2013,Bewerunge2016,Baeuerle2018,Volpe2015,Nishizawa2017} or, more naturally, 
may arise in autochemotactic particles, i.e., if the particles themselves are part of the production mechanism of the chemical 
substance they react to. In particular, examples of the latter include biological systems such as bacteria 
\cite{Adler1966, Keller1971} and army ants \cite{Couzin2003}, as well as synthetic microswimmers such as active colloidal particles 
\cite{Bechinger2016,Liebchen2015,Liebchen2016,Liebchen2017,Saha2014,Sengupta2009,Sengupta2011} or self-propelling droplets 
\cite{Jin2017,Jin2018}.

In the context of many-particle systems, the topic of pattern formation \cite{Cross2009,CrossHohenberg1993,Donella1997,Zakine2018,Chacko2015,Zimmermann2013,Lushi_2018} is of central interest.
In particular, the Ginzburg-Landau \cite{Aronson2002,Puzyrev2014,Ciszak2015, Emmerich2012} and Swift-Hohenberg equations \cite{SwiftHohenberg1977} 
are widely used to study pattern formation. Most of these studies present a coarse-grained treatment using effective continuum theories but do not resolve the individual particles.
One pattern which is commonly observed in many different systems is traveling waves or moving bands of particles.
Examples include actin-waves formed in the biological actin-myosin systems \cite{Gerisch2004,LeGoff2016,Schaller2010}, metachronal waves
in cilia arrays \cite{metachronal_Uchida},
the patterning in systems of active agents under various settings \cite{Vicsek1995,Chate2008,Menzel2013,Zimmermann2013,Ophaus2018,Reinken2018,Schnyder2017}, 
the formation of bands in passive colloidal 
suspensions driven by ac \cite{Vissers2011,Wysocki2009} or dc 
\cite{Dzubiella2002, Reichhardt2007, Ikeda2012, Reichhardt2018,ReichhardsSM2018,Dzubiella2002JPCM} fields, and phase separating mixtures 
\cite{Okuzono2001,Okuzono2003,Zimmermann2013}. Recent work on pattern forming systems also considers the effect of time-delayed feedback using continuum theories \cite{Gurevich2013PRE,Tabbert2017,Gurevich2013,Ciszak2015}.

In this paper, we present a study of feedback-driven colloidal particles as an example of a feedback system 
of discrete components considered on the fundamental particle level. In our model, the particles are subjected to a feedback potential 
driving them away from their previous positions. Using Brownian dynamics computer simulations and 
dynamical density functional theory \cite{Marconi1999,Marconi2000,Archer2004,Espanol2009,Evans2016,Wittkowski2011}, we show that this repulsive feedback leads to self-organization of the particles into a moving band structure reminiscent of a traveling wave.
Remarkably, this ordering takes place despite the absence of any attractive interactions in the system, for which static band formation is known to occur \cite{OlsonReichhard2010,Zhao2012}.
The width of the bands and their
propagation speed can be tuned by the delay time and the range of the imposed repulsive delay potential. Finally, we demonstrate that
traveling band formation also persists under strong confinement leading, in circularly confined systems, to globally rotating and spiraling bands \cite{Wioland2013}.

Our model can be realized in experiments for colloidal suspensions. The suspensions can be exposed to a potential energy landscape using optical fields which are programmed via a feedback loop \cite{Baeuerle2018,Leyman2018,Lavergne2019}. 
Typically the colloids are attracted towards the intensity maximum of the optical field. However, by inverting the intensity landscape a repulsion is achieved, in which case the particles are driven away from the dark regions.

The paper is organized as follows: In the following section we introduce the underlying Langevin equation including a delay term \cite{Ohira2000,Ohira2005,Ohira2009,Trimper2004,LeBerre1990,Guillouzic1999,Atay2010,Loos2019} describing the dynamics of the system. 
We continue with presenting our simulation results in Sec.~\ref{sec_simu}. A prediction of the observed traveling wave formation is derived from 
dynamic density functional theory in Sec.~\ref{sec_DDFT}. Subsequently, we consider confinement effects in Sec.~\ref{sec_confine}. Finally, we conclude with a summary of our main findings and an outlook 
to possible extensions of the system in Sec.~\ref{sec_conclusion}.

\section{Model and Brownian dynamics computer simulations}
The Brownian dynamics of $N$ colloidal particles in two spatial dimensions is described 
by their time-dependent positions $\vect{r}_i(t)$ ($i=1,\ldots,N$) and governed by the following Langevin equation
\begin{align}
\gamma\frac{\dif\vect{r}_i}{\dif t}=\vect{f}_i(t)
&+\sum_{j=1}^{N}\vect{F}\left(\vect{r}_i(t)-\vect{r}_{j}(t-\tau)\right)\nonumber\\
&+\sum_{\substack{j=1\\ j\ne i}}^{N}\vect{F}_\mathrm{Yuk}\left(\vect{r}_i(t)-\vect{r}_j(t)\right)\,.\label{eq_eom_manyp}
\end{align}
which can be viewed as a force balance equation. The left-hand side of Eq.~(\ref{eq_eom_manyp}) contains the Stokes drag force with
$\gamma$ denoting the friction coefficient. The Gaussian random force $\vect{f}_i(t)$ mimics the collision of the particle with solvent molecules.
This stochastic force is characterized by its first 
two moments $\langle\vect{f}_i(t)\rangle=0$ and $\langle \vect{f}_i(t)\otimes\vect{f}_j(t')\rangle=2D\gamma^2\mathds{1}\delta\left(t-t'\right)\delta_{ij}$, where 
$D$ is the short-time diffusion coefficient of the particles, $\delta\left(t\right)$ is the Dirac delta function and $\delta_{ij}$ denotes the Kronecker delta.
The important new ingredients in Eq.~(\ref{eq_eom_manyp}) are the feedback forces $\vect{F}\left(\vect{r}_i(t)-\vect{r}_{j}(t-\tau)\right)$.
These forces are evaluated at distances between the actual position $\vect{r}_i(t)$ of particle $i$ and the \textit{former}
positions $\vect{r}_j(t-\tau)$ of the other particles $j$ (where the special case $i=j$ is included).
Here, $\tau$ is the time difference, which we refer to as the \textit{delay time} of the feedback. 

We derive $\vect{F}(\vect{r})$ from a potential $\Vext(r)$ 
as 
\begin{align}
\vect{F}(\vect{r})=-\nabla_\vecr \Vext\left(r\right)
\end{align} and assume for simplicity a Gaussian form
\begin{align}
\Vext(r)=A\ \exp{(-\frac{r^2}{2b^2})}\,,\label{eq_Vext}
\end{align}
characterized by an energy amplitude $A$ and a range $b$. The Gaussian potential form is a good approximation for optical systems such as optical tweezers and occurs naturally for autochemotactic particles \cite{Sengupta2009}. Here, we confine ourselves to the case of repulsive feedback potentials such that the energy amplitude $A>0$ is positive and the special case $A=0$ serves as an equilibrium reference case.
For $A>0$, all particles are driven away from the past positions of all particles including their own.

Finally, the equations of motion include direct particle-particle interaction forces
\begin{equation}
\vect{F}_\mathrm{Yuk}(\vect{r})=-\nabla_\vecr \VYu(r)
\end{equation}
via a repulsive Yukawa pair potential
\begin{equation}
\VYu(r)=\frac{V_0}{r}\ \exp{(-\kappa r)}
\end{equation}
involving an inverse range $\kappa$ and an amplitude $V_0$.

We perform Brownian dynamics simulations with a square simulation box of length $L$ and periodic 
boundary conditions with $N=6400$ particles. Some of the simulations were repeated in a rectangular box, in order to obtain stable traveling bands. This was necessary because at the onset of the formation of traveling bands, the band stability is highly dependent on the commensurability of the box size and the preferred wavelength. Possible wavelengths in the finite system are restricted to those being commensurate with the periodic boundaries, which thus requires the system length to be adjusted. The equation for the particle positions, Eq.~(\ref{eq_eom_manyp}), is integrated using an explicit Euler scheme with a finite
time step of $\Delta t=10^{-4}\tau_0$, where $\tau_0=b^2/D$ denotes the Brownian time scale.

In the following, lengths are normalized to the feedback potential range $b$ and times to the Brownian time $\tau_0=b^2/D$. 
Energies are given in terms of the thermal energy $k_BT\equiv D\gamma$.
In order to keep the set of parameters limited, we maintain $V_0=60\,bk_BT$, $\kappa=4.5/b$ and $\tau=0.25\,\tau_0$ constant in our units. Furthermore, we use a number density $\rho_0=N/L^2=1/b^2$. The units are dropped hereafter for ease of notation. We use the feedback amplitude $A$ as a control parameter and investigate the change in the system structure and dynamics as a function of it.

Our simulation protocol is as follows: First, the system is equilibrated without any feedback 
potential (corresponding to a two-dimensional pure Yukawa system \cite{HL_JPCM}), after which the positions are 
recorded for updating the feedback potential. We define $t=0$ as the 
time at which the feedback potential is first introduced into the system. Subsequently, the relaxation of the system is monitored for a long time, several hundreds of time units.

\section{Simulation Results}\label{sec_simu}
\subsection{Band formation}
\begin{figure}[tb]
\includegraphics[width=0.99\columnwidth]{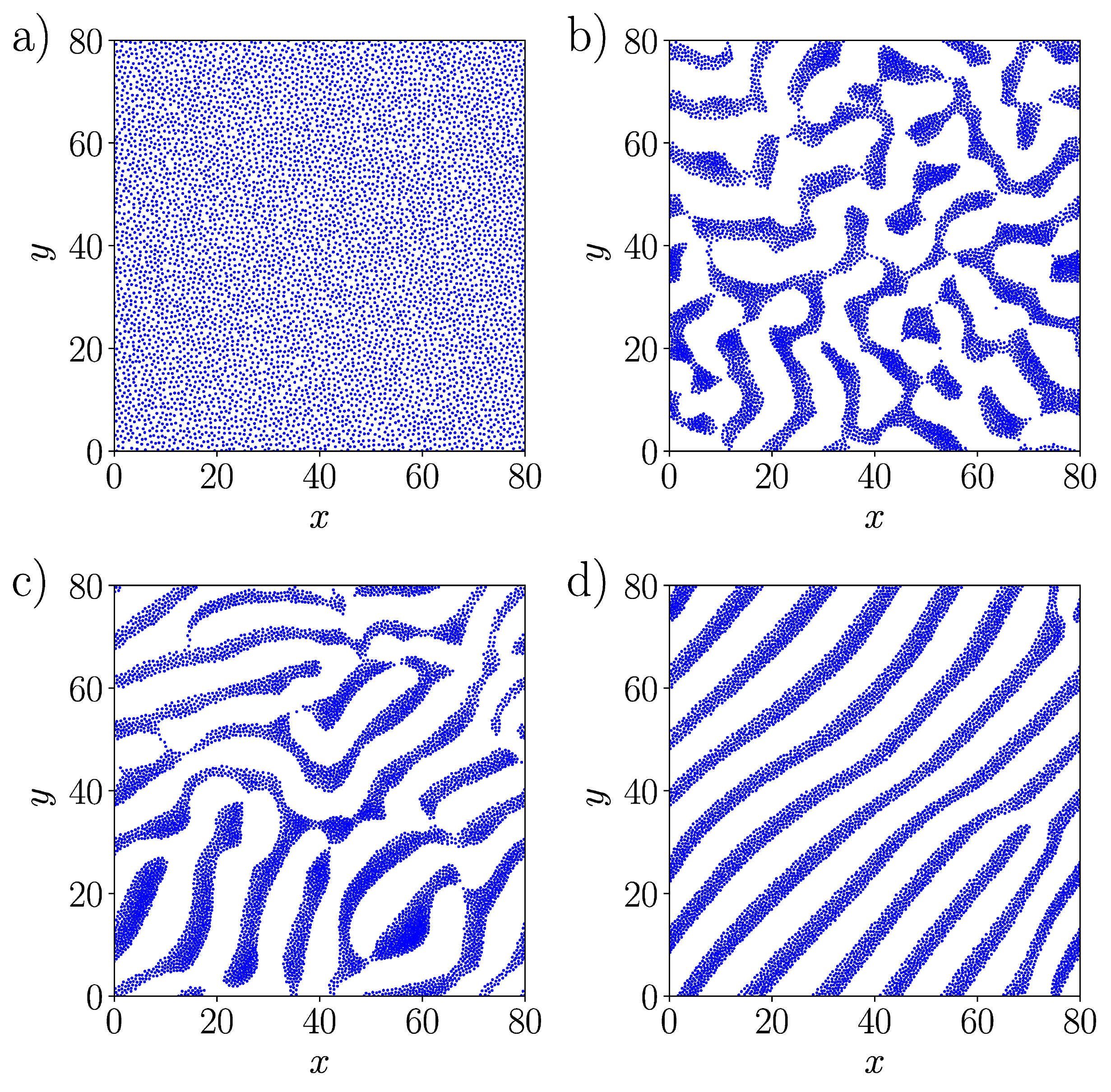}
\caption{Formation of bands. The plots show snapshots of the system at times (a) $t=0$, (b) $t=10$, (c) $t=50$, and (d) $t=100$ for the case of a strong feedback potential ($A=20$). Separation into crowded and empty regions is followed by the formation of bands.\label{fig_band_formation}}
\end{figure}

Figure \ref{fig_band_formation} shows a typical instance of system relaxation in the case of strong feedback potentials. The initial equilibrated homogeneous fluid state
in Fig.~\ref{fig_band_formation}(a) spontaneously separates into two regions which are either empty or crowded (i.e.,~exhibiting a high density of particles). This demixed state coarsens further as a function of time into a configuration of system-spanning straight bands at long times, resembling those observed in mixtures of particles subjected to bidisperse Magnus forces \cite{OlsonReichhardt2019}. The width of the formed bands depends on the specific parameters of the particle repulsion and the feedback potential as well as the delay time $\tau$. In particular, no band formation is observed for very small delay times, in which case the feedback potential can be understood as modified direct interactions between the particles, while larger delay times lead to an increase in the time scale on which band formation takes place.
The orientation of the bands is tilted relative to the quadratic box with the angle depending on the initial configuration as well as the commensurability of the wavelength and the box size. The commensurability condition leads to a finite set of possible orientations for the bands while the initialization determines which of these is realized.

The emerging bands are observed to move collectively along the normal of their interfaces.
The empty regions are found at the former particle positions, i.e., the positions where the potential is inserted, suggesting that particles try to effectively 
avoid regions where they have 
been a time $\tau$ before. In more detail, the occurrence of the separation into empty and crowded regions can be explained in the following way:
\begin{figure}[tbh]
\includegraphics[width=0.99\columnwidth]{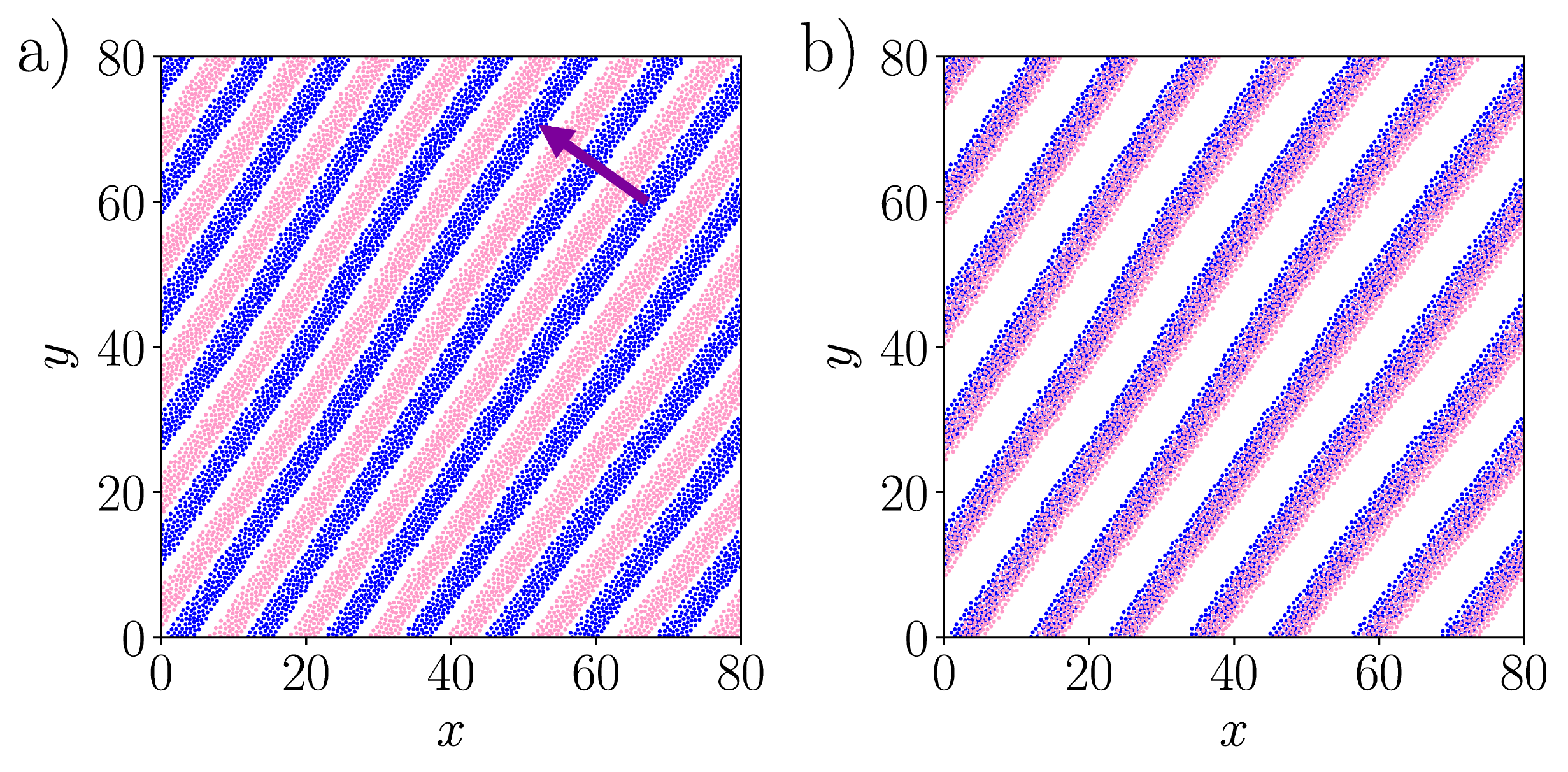}
\caption{Particle positions at times $t=500$ (blue) and (a) $t=500.25$ and (b) $t=500.5$ (pink) for feedback amplitude $A=20$. The particles move approximately one bandwidth in the normal direction within a feedback time $\tau=0.25$. The direction of movement is indicated by the arrow in (a).\label{fig_band_tau}}
\end{figure} 
If in a disordered system as in Fig.~\ref{fig_band_formation}(a) there are by chance more particles at a particular position at time $t$, 
the strong repulsive potential imposed at time $t+\tau$ leads to fewer particles at this position, which in turn leads to a small potential and more particles 
at time $t+2\tau$. The feedback potential thus leads to a self-ordering and the particle distribution effectively changes between one state 
and its negative image with period $2\tau$.
The easiest way to achieve this, namely, the one with the fewest collisions between the particles, is a collective movement into one direction 
as given for the moving lamellar phase seen in Fig.~\ref{fig_band_tau}.

Based on the previous consideration, 
a band moves over its full periodicity $\lambda_\mathrm{b}$ during twice the delay time $\tau$
such that a scaling expression for the magnitude of the expected band velocity is obtained as
\begin{equation}
v_\mathrm{s} = \frac{\lambda_\mathrm{b}}{2\tau}\,.\label{eq_vtheo}
\end{equation}
Remarkably, knowledge of the static property, namely, the wavelength and orientation of the band structure, thus provides an estimate of the dynamics of the system, i.e., the band velocity. Likewise, determining the velocity of the bands for a given feedback time yields an approximative value for the band periodicity $\lambda_\mathrm{b}$. Moreover, through determination of the velocity and periodicity, the feedback time can be estimated, which, in particular in biological systems, might not be easily accessible otherwise.

The prediction for the band velocity obtained from the scaling expression can be compared to the simulation results. For the latter case, we 
use the systematic force
\begin{align}
\vect{F}_i(t)=\Bigg[&\sum_{j=1}^{N}\vect{F}\left(\vect{r}_i(t)-\vect{r}_{j}(t-\tau)\right)
\nonumber\\&+\sum_{\substack{j=1\\ j\ne i}}^{N}\vect{F}_\mathrm{Yuk}\left(\vect{r}_i(t)-\vect{r}_j(t)\right)\Bigg]\,.\label{eq_FD}
\end{align}
acting on particle $i$ at time $t$ to define an instantaneous drift velocity
\begin{align}
\vect{v}_i(t) = \frac{\vect{F}_i(t)}{\gamma}\label{eq_vD}
\end{align}
of this particle.
From this expression, the mean global drift velocity $\vect{v}$ is then obtained by averaging over all particles as
\begin{align}
\vect{v} &=\frac{1}{N} \sum_{i=1}^{N} \langle\vect{v}_i(t')\rangle\,,\label{eq_vbar_D}
\end{align}
where
\begin{align}
\langle B(t')\rangle =  \frac{1}{T} \int_{t_0}^{t_0+T} \dif t' B(t') \label{average}
\end{align}
denotes an average taken over time $t'$ for an observable $B(t')$. The time $t_0$ is bigger than a typical relaxation time of the system and $T$ is the width of the time window over which the average is performed. Here, we use $t_0=500$ and $T=500$. In the ordered band state the mean drift velocity equals the velocity of the bands. The comparison between the two velocities is shown in the next section as a function of the feedback potential amplitude $A$ [see Fig.~\ref{fig_v_theo_simu}].

\FloatBarrier
\subsection{Dependence on the feedback strength}
\begin{figure}[htb]
\includegraphics[width=0.99\columnwidth]{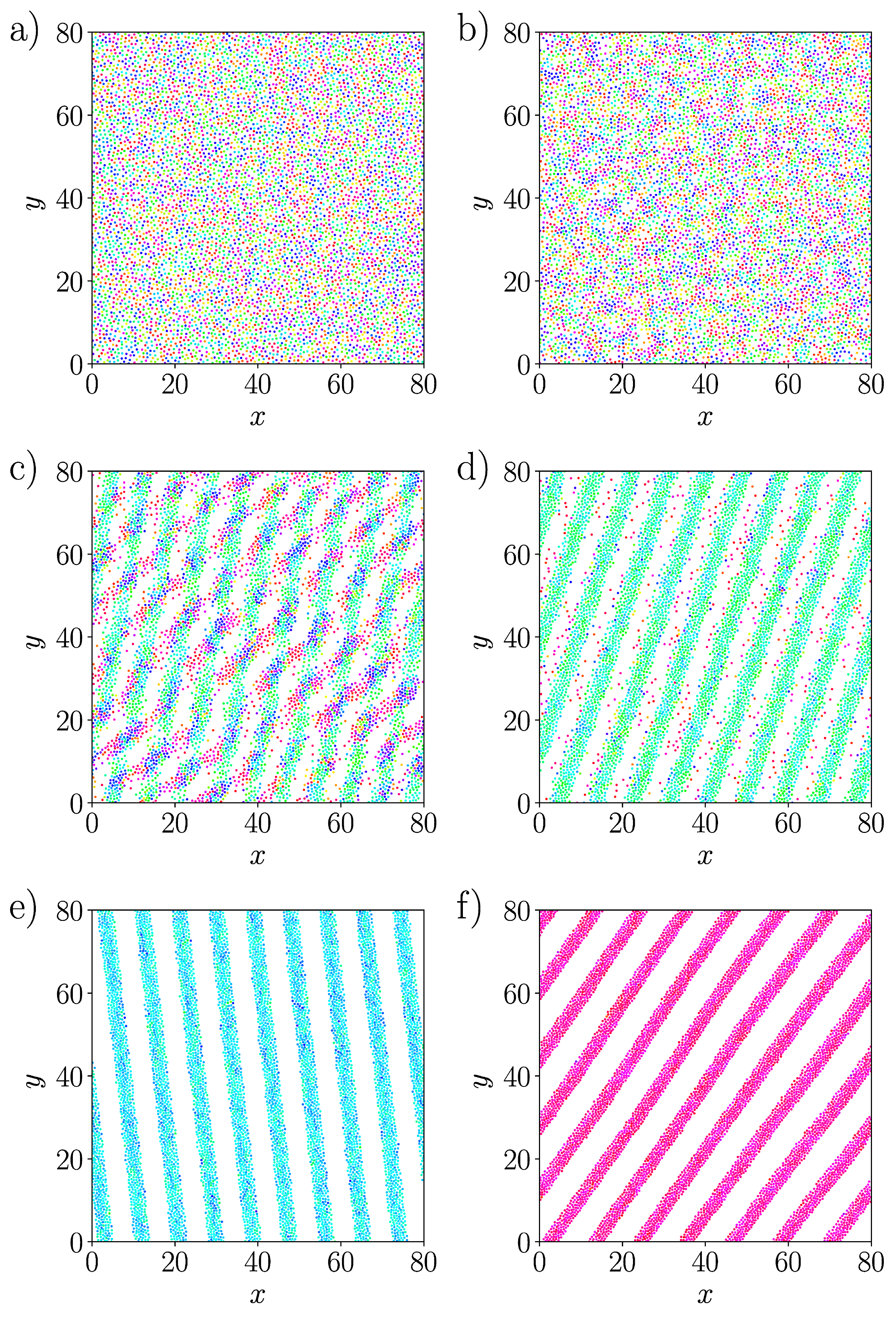}
\flushright
\includegraphics[width=0.6\columnwidth]{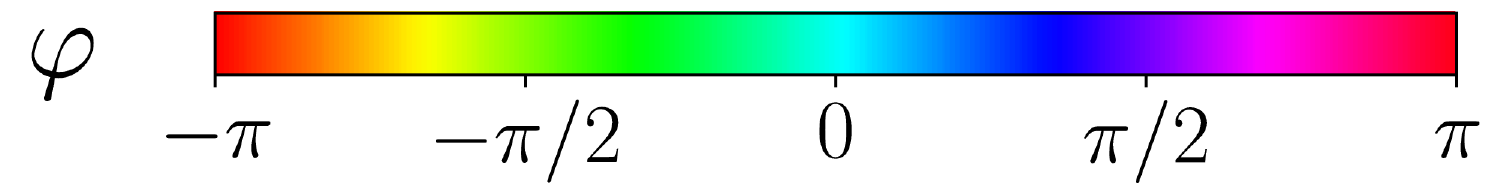}
\caption{System snapshots for feedback potential strengths (a) $A=1$, (b) $A=5$, (c) $A=6$, (d) $A=6.5$, (e) $A=10$, and (f) $A=20$ after $t=500$. 
A band pattern is formed initially for sufficiently high strengths of the feedback potential. The direction of motion of the particles 
is extracted from the individual drift velocities $\vect{v}_i(t)=|\vect{v}_i|\left(\cos\varphi_i,\sin\varphi_i\right)$ and the angle $\varphi_i$ is indicated by color. \label{fig_snapshot_t500}}
\end{figure}

The system shows different structuring depending on the amplitude of the applied feedback potential $A$. The patterns are shown in Fig.~\ref{fig_snapshot_t500} with the color code indicating the drift velocity directions of the individual particles. As a general result, we find that the average of these velocities, the mean drift velocity $\vect{v}$, is typically normal to the band direction.

With respect to potential strength $A$, we observe that, while for small $A\lesssim 1$ [Fig.~\ref{fig_snapshot_t500}(a)], diffusion prevents any structure formation, higher potential amplitudes lead to patterning [Fig.~\ref{fig_snapshot_t500}(b)] and, for even larger potential strength, to the formation of a band structure. The necessary potential amplitude for pattern formation can be estimated by considering at which point the feedback force $\vect{F}$ becomes comparable to the interparticle repulsion force $\vect{F}_\mathrm{Yuk}$.
From the condition $\vect{F}^2=\vect{F}_\mathrm{Yuk}^2$, we find that $A\approx 6$, which is in reasonable agreement with the simulation results which indicate the start of band formation at $5\le A^*\le6$. Close to this threshold, the band state can be constituted of two distinct band orientations [Fig.~\ref{fig_snapshot_t500}(c)]. Further increasing the potential leads to a single band orientation with a considerable number of particles traveling in the opposite direction [Fig.~\ref{fig_snapshot_t500}(d)]. At even higher potentials [Figs.~\ref{fig_snapshot_t500}(e) and \ref{fig_snapshot_t500}(f)], the system forms stable bands with all particles moving as part of the bands into the same direction.

We have checked the system for hysteretic behavior via additional simulations. For a finite system, a small hysteresis effect is observed in forming the bands, which appears to be due to the finite system size and the prescribed periodic boundary conditions.

\begin{figure}[htb]
\includegraphics[width=0.7\columnwidth]{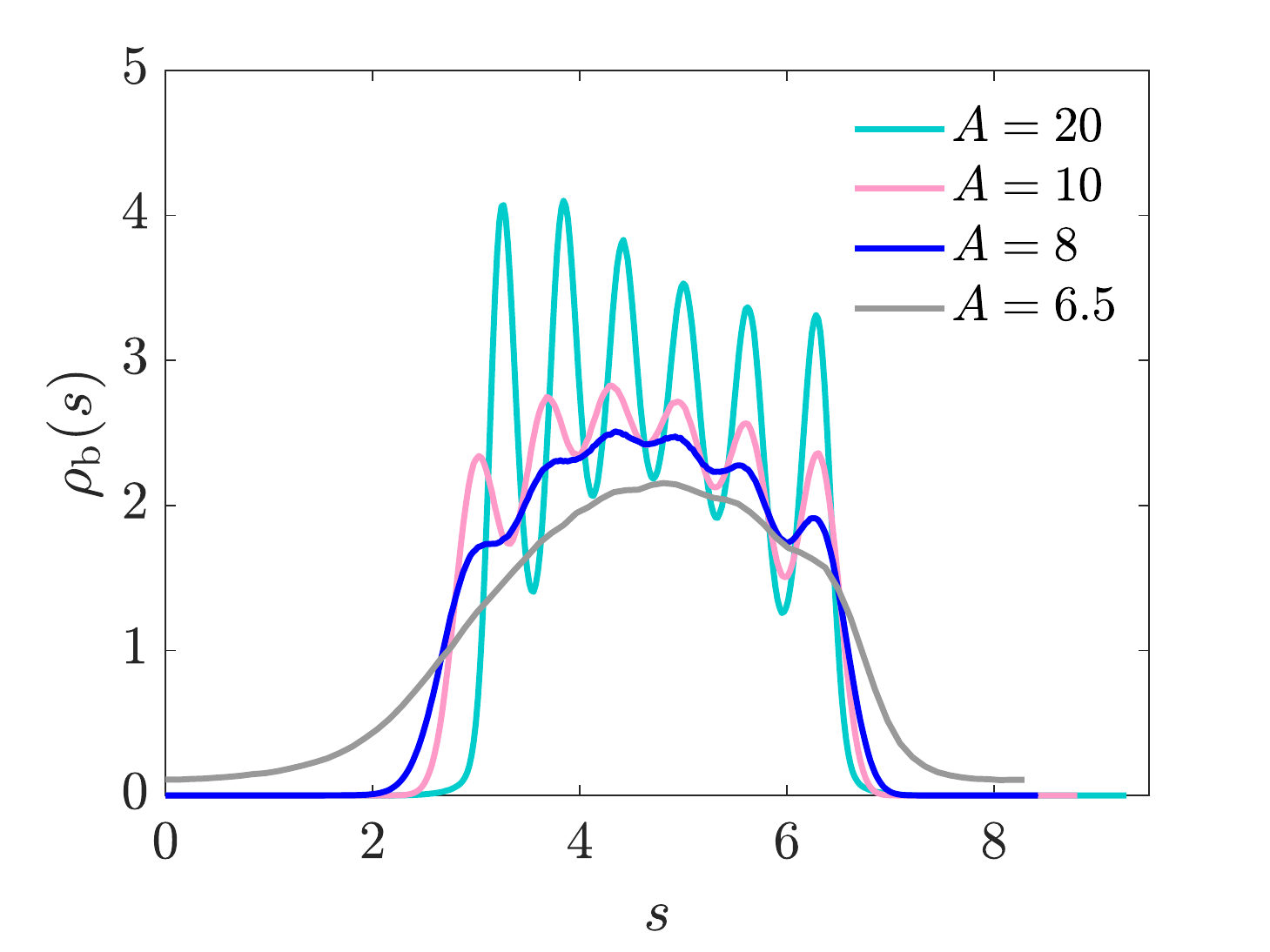}
\caption{Density profile $\rho_\mathrm{b}(s)$ of the bands in the co-moving frame, as a function of the position $s$ in the band drift direction for different feedback potential strengths $A$.}\label{fig_dens_profile}
\end{figure}

The density distribution within the bands $\rho_\mathrm{b}$ can be determined by changing to a co-moving frame. Taking an average along the band tangential, the density only depends on the position $s$ in the drift direction $\hat{\mathbf{e}}_\mathrm{v}$,
\begin{equation}
\rho_\mathrm{b}\left(s\right)= \Big\langle \sum_{i=1}^N \delta\left(s-\vect{r}_i(t')\cdot\hat{\mathbf{e}}_\mathrm{v} \right)\Big\rangle\,.
\end{equation}
One period of these profiles is shown in Fig.~\ref{fig_dens_profile}, revealing an increasing layering of particles for strong feedback amplitudes and hence an increasing internal order of the bands.

\FloatBarrier

\begin{figure}[htb]
\includegraphics[width=0.99\columnwidth]{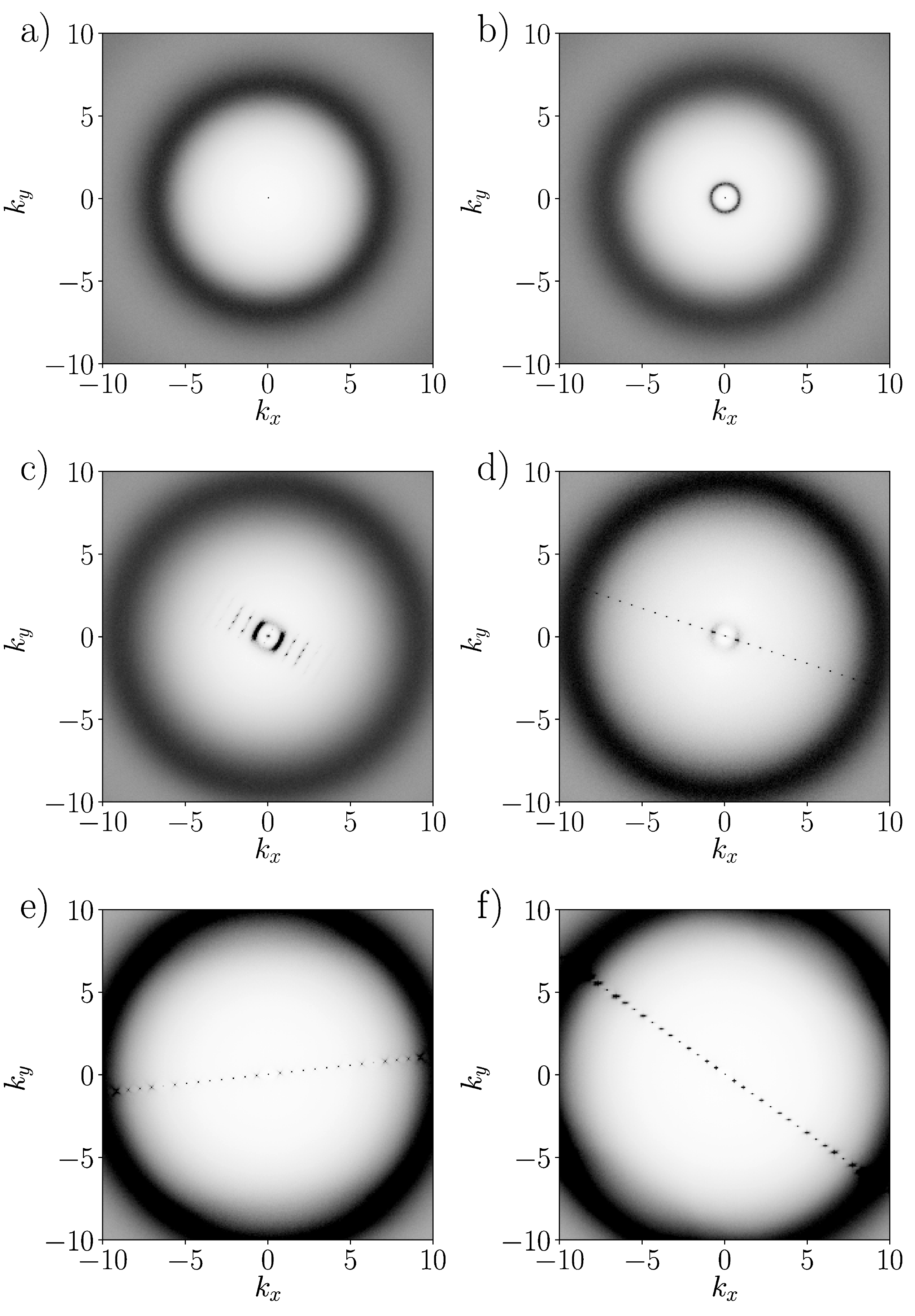}
\flushright
\includegraphics[width=0.6\columnwidth]{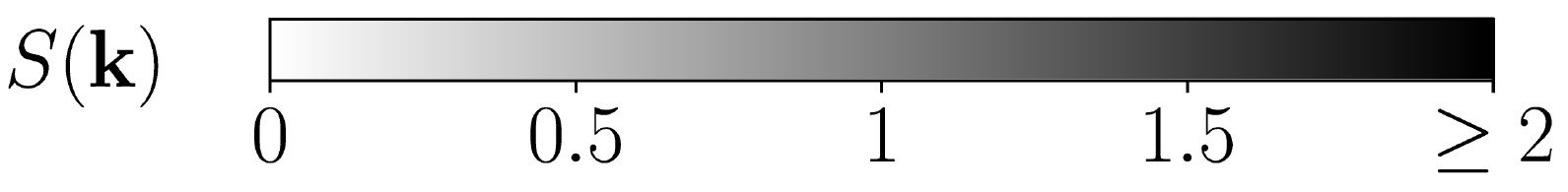}
\caption{Two-dimensional structure factor $S(\vect{k})$ for the systems of Fig.~\ref{fig_snapshot_t500}, i.e., for feedback amplitude (a) $A=1$, (b) $A=5$, (c) $A=6$, (d) $A=6.5$, (e) $A=10$, and (f) $A=20$. The outer black ring corresponds to the mean particle distance, the inner black ring, visible in (b)\textendash(d), to patterning at the wave number $k^*$. The system transitions from directionally homogeneous patterning at wave number $k^*$ to preferred orientations between $A=5$ and $6$.}\label{fig_Skxky}
\end{figure}

In the following, the formed patterns are explored in more detail via the structure factor, defined by
\begin{equation}
S(\vect{k})=\Bigg\langle\frac{1}{N}\sum_{i,j=1}^{N}\rme^{-i\vect{k}\left(\vect{r}_i(t')-\vect{r}_j(t')\right)}\Bigg\rangle\,.
\end{equation} 
Figure~\ref{fig_Skxky} shows $S(\vect{k})$ for the cases presented in Fig.~\ref{fig_snapshot_t500} as a function of the components of the wave vector $\vect{k}=\left(k_x,k_y\right)$. The outer black ring corresponds to the mean particle distance. For higher potential, its radius is shifted away from the equilibrium value $2\pi\rho_0^{1/2}$ towards higher wave numbers, 
indicating that the mean distance between particles in the band structure is considerably smaller than the one in the homogeneous system without the feedback potential.

\begin{figure}[htb]
\includegraphics[width=0.7\columnwidth]{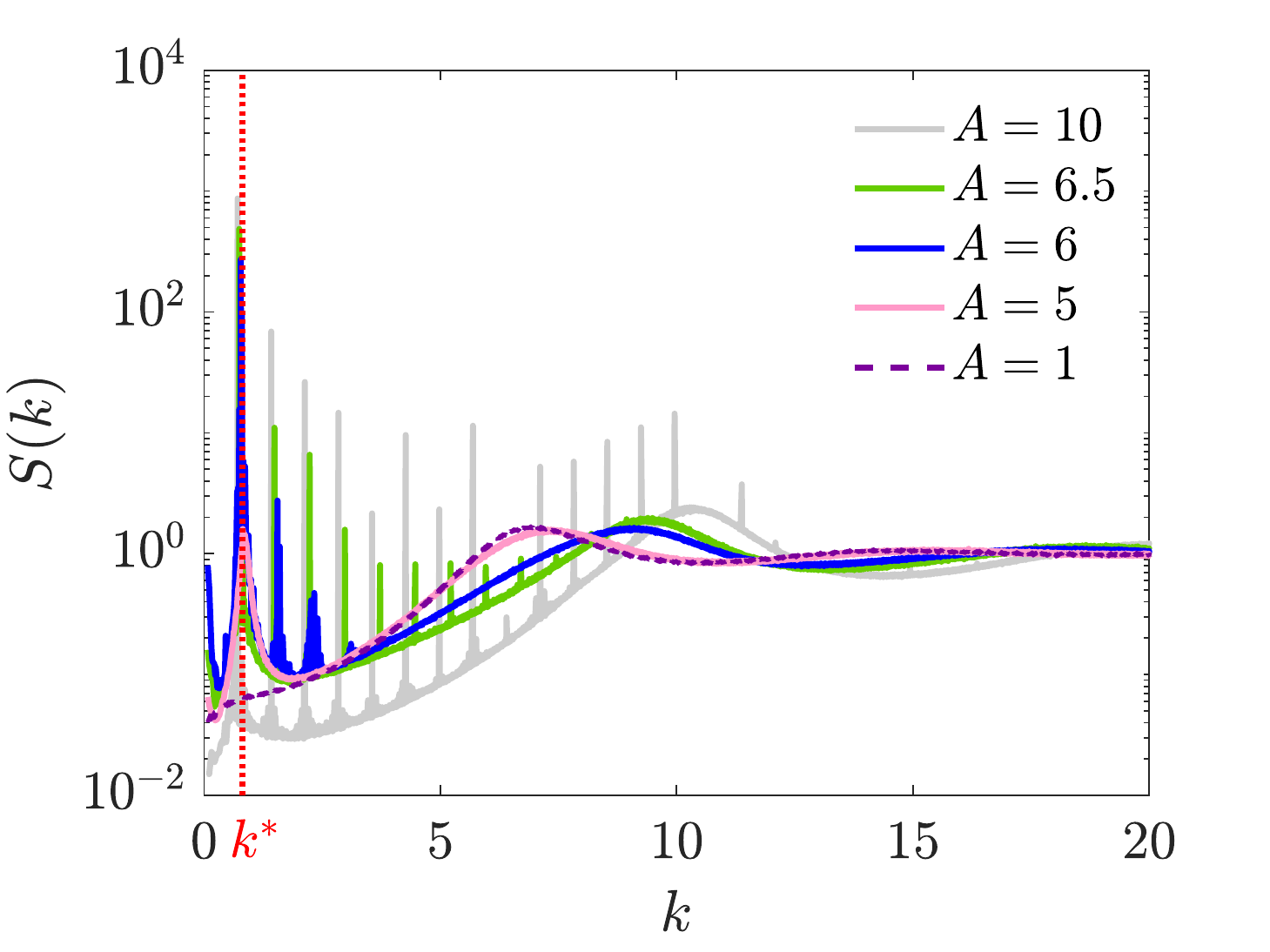}\hfill
\caption{Azimuthally averaged structure factor $S(k)$ for different values of the feedback potential strength $A$. The structure factor shows the appearance of a new structure of wave number $k^*\approx0.82$ in the system, corresponding to the wavelength of the traveling bands. For stronger potentials  additional peaks appear at multiples of this wave number indicating higher harmonics.}\label{fig_S}
\end{figure}

Further, the inner black ring first appearing in Fig.~\ref{fig_Skxky}(b) represents the ordering due to the feedback potential. 
With respect to the feedback potential strength $A$, different stages of ordering are observed at this wave number which we denote by $k^*$. First, for small potentials, directionally independent patterning at $k^* \approx 0.82$ is found [Fig.~\ref{fig_Skxky}(b)], which 
for stronger amplitudes develops a directional dependence with two preferred band orientations [Fig.~\ref{fig_Skxky}(c)]. While medium potentials [Figs.~\ref{fig_Skxky}(c) and \ref{fig_Skxky}(d)] still show some remainder of the initial orientationally-independent ordering, indicated by the light gray ring at $k^*$, this feature disappears at large feedback potential strength $A$ [Figs.~\ref{fig_Skxky}(e) and \ref{fig_Skxky}(f)]. For strong feedback $A$, only a single band orientation is found.  For large potential amplitudes [Figs.~\ref{fig_Skxky}(d)\textendash(f)], the nonsinusoidal form of the density profile, visualized in Fig.~\ref{fig_dens_profile}, is reflected in higher harmonics which lead to additional peaks at multiples of $k^*$ in the structure factor.
For easier comparison, the azimuthal average $S(k)$ is shown in Fig.~\ref{fig_S} for different feedback strengths $A$, illustrating the shift in the mean particle distance as well as the growth of the new wave number $k^*$ and its higher harmonics.

\begin{figure}[htb]
\includegraphics[width=0.7\columnwidth]{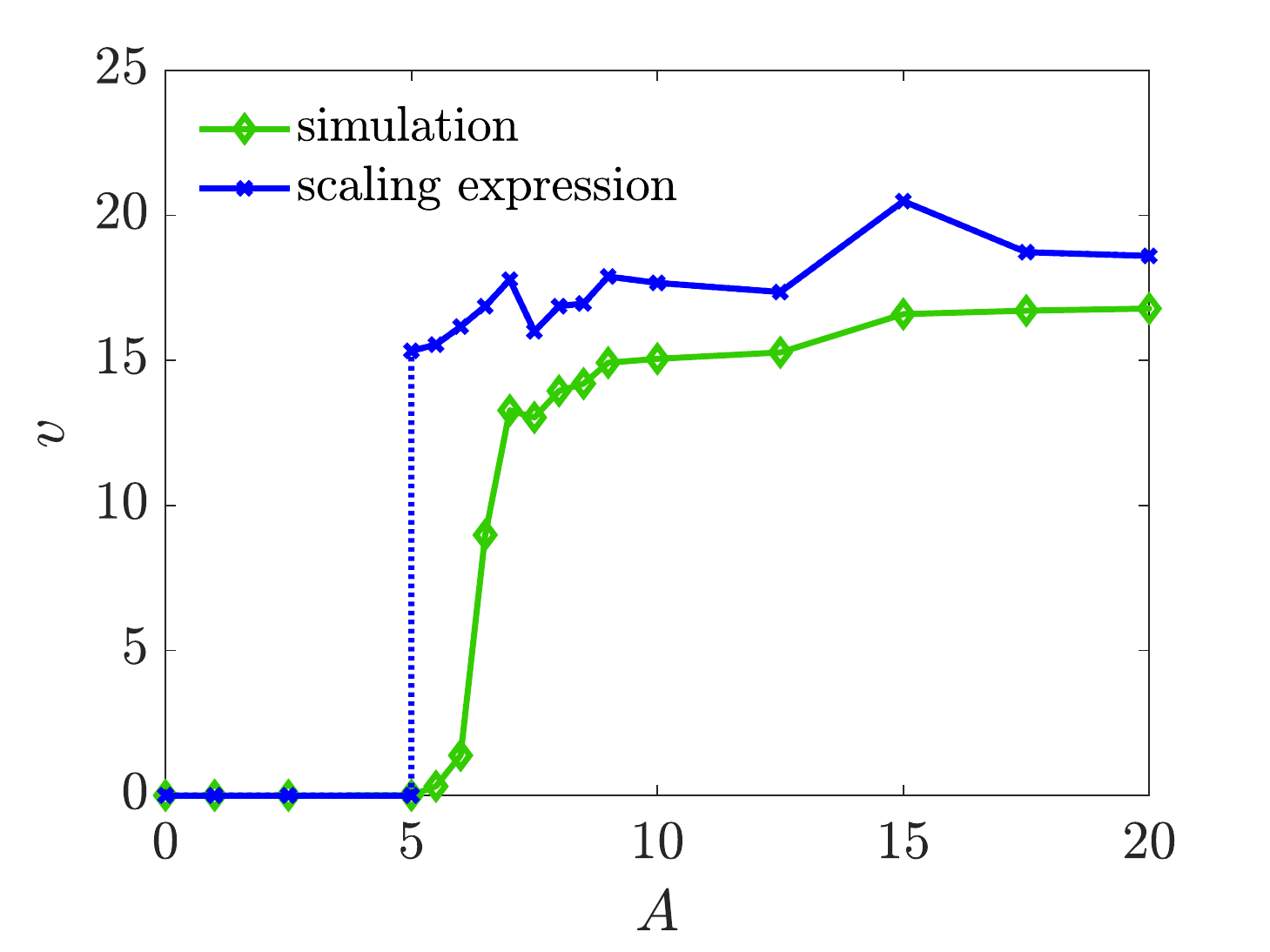}
\caption{Mean drift velocity obtained from simulations [Eq.~(\ref{eq_vbar_D})] compared to the band velocity predicted via the scaling expression [Eq.~(\ref{eq_vtheo})]. For the latter, the wavelength $\lambda_\mathrm{b}=2\pi/k^*$ is a necessary input; its value is extracted from the structure factor obtained through simulations. \label{fig_v_theo_simu}}
\end{figure}

Extracting the wavelength of the band pattern from the structure factor $S(k)$, allows for a comparison between the estimated value of the band velocity obtained from the scaling expression introduced in the previous section and the simulation results.
The prediction for the band velocity according to the scaling expression $v_\mathrm{s}$, given by Eq.~(\ref{eq_vtheo}), and the simulation results for the magnitude $v$ of the global drift velocity $\vect{v}$, defined by Eq.~(\ref{eq_vbar_D}), are shown in Fig.~\ref{fig_v_theo_simu}. Above a threshold value $5\le A^*\le6$, $v$ increases sharply. For high feedback amplitudes $A$ it saturates to a value that depends on the specific choice of the remaining potential parameters as well as the delay time $\tau$. The agreement between the theoretical prediction and the simulation results is acceptable but not exact. The reason for the discrepancy is that the bands filled with particles and the empty spaces in between are not exactly equal in width, a consequence of the feedback force pushing the particles from behind. The difference in width is also visible in Fig.~\ref{fig_band_tau}. The length traveled by the bands is thus overestimated when using half the wavelength of the band structure instead of the width of the bands, leading to a slightly higher predicted velocity for the present parameters.

\begin{figure}[htb]
\includegraphics[width=0.7\columnwidth]{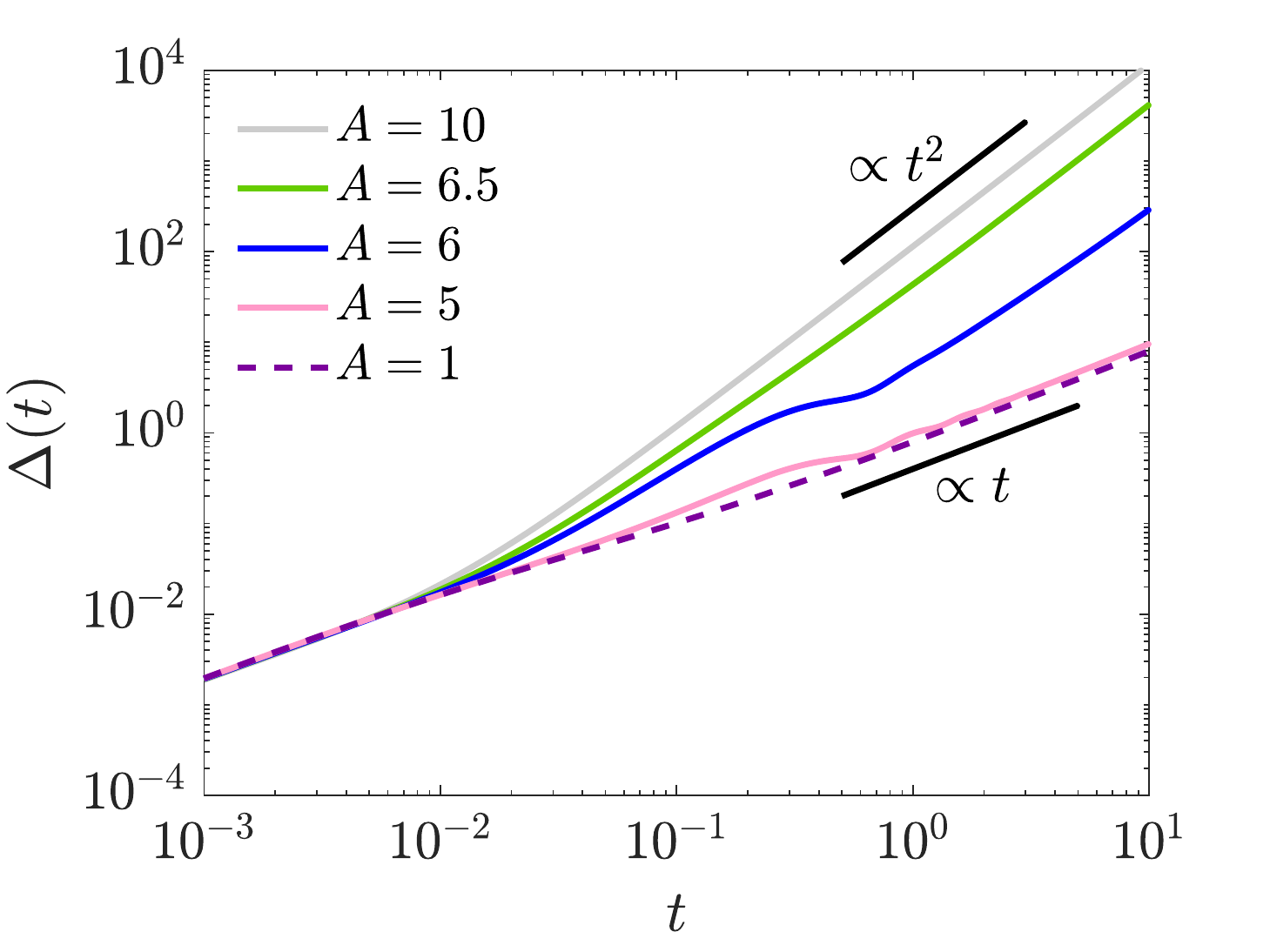}
\caption{MSD $\Delta(t)$ for different values of the feedback potential strength $A$. Within the time window considered, the long-time MSD changes from diffusive to ballistic behavior.}\label{fig_MSD}
\end{figure}

For the individual particles, the directed drift motion becomes visible in the mean-squared displacement (MSD), which we define by
\begin{equation}
\Delta(t)=\Bigg\langle \frac{1}{N} \sum_{i=1}^{N} \left( \vect{r}_i\left(t'+t\right) -\vect{r}_i\left(t'\right)\right)^2  \Bigg\rangle\,.
\end{equation} 
The MSD changes qualitatively with the onset of the traveling wave instability: Increasing $A$ over the threshold value for band formation changes the long time behavior from diffusive ($\propto t$) to a directed drift
motion ($\propto t^2$), similar to what is found for active (self-propelled) Brownian particles \cite{GolestanianPRL2017,Bechinger2016} (see Fig.~\ref{fig_MSD}).
Thus, the feedback potential effectively provides a source of self-propulsion.

\FloatBarrier

\section{Dynamical density functional theory}\label{sec_DDFT}
The previously observed values for the potential strength $A^*$ and wave number $k^*$ characterizing the onset of the pattern formation instability
are now derived from microscopic dynamic density functional theory (DDFT). The theory requires the particle interactions as the only input.

DDFT is based on the Smoluchowski equation \cite{Dhont1996} corresponding to the Langevin equation~(\ref{eq_eom_manyp}). The equivalent time-delayed Smoluchowski equation is known for the one-particle case \cite{Frank2005,Frank2005_PRE72,Loos2017,Loos2019,Guillouzic1999,Atay2010} and is briefly discussed in Sec.~\ref{sec_DDFT} A. Subsequently, this single-particle case is extended to the case of many particles (in Sec.~\ref{sec_DDFT} B) and the approximations used to obtain a self-contained DDFT equation are introduced in Sec.~\ref{sec_DDFT} C. Finally, the system's stability against traveling-waves is investigated within DDFT in Sec.~\ref{sec_DDFT} D and the obtained predictions are compared to the simulation results in Sec.~\ref{sec_DDFT} E.

\subsection{One-particle time-delayed Smoluchowski equation}
To determine the time-delayed Smoluchowski equation equivalent to Eq.~(\ref{eq_eom_manyp}), we first revisit the case of just a single feedback-driven particle. In \cite{Frank2005}, the Smoluchowski equation corresponding to a time-delayed Langevin equation was derived. Equations~(1), (12), and (14) of this work are relevant here and are reproduced below for the case of a constant noise amplitude. Specifically, in \cite{Frank2005} it was shown that a time-delayed Langevin equation of the form 
\begin{equation}
\frac{\partial}{\partial t} X(t)=h\left(X(t),X(t-\tau)\right)+\sqrt{2D}\ \Gamma(t)\,,\label{eq_Frank}
\end{equation}
for a general state variable $X(t)$ subject to a Gaussian noise $\Gamma(t)$, defined by $\langle\Gamma(t)\rangle=0$ and $\langle\Gamma(t)\Gamma(t')\rangle=\delta(t-t')$, is equivalent to the Smoluchowski equation
\begin{equation}
\frac{\partial}{\partial t}w(x,t|x_\tau,t_\tau)\Big|_{t_\tau=t-\tau}=\hat{L}(x,\nabla_x,x_\tau)w(x,t|x_\tau,t_\tau)\label{eq_Smolu_Frank}
\end{equation}
with 
\begin{align}
\hat{L}(x,\nabla_x,x_\tau)=&-\frac{\partial}{\partial x}h(x,x_\tau)+D\frac{\partial^2}{\partial x^2}\,.
\end{align}
Here, the conditional probability $w(x,t|x_\tau,t_\tau)$ gives the probability for the system to be in state $x$ at time $t$, under the condition that it was in state $x_\tau$ at time $t_\tau=t-\tau$. The time derivative on the left-hand side of Eq.~(\ref{eq_Smolu_Frank}) only acts on $t$ but not $t_\tau$. The one-particle equivalent of our Eq.~(\ref{eq_eom_manyp}) is contained in this solution by identifying the actual and the time-shifted system states $x$ and $x_\tau$ with the particle positions $\vect{r}$ and $\vect{r}_\tau$. We set $h(\vect{r},\vect{r}_\tau)=F(\vect{r},\vect{r}_\tau)/\gamma$ for which 
\begin{align}
\hat{L}(\vect{r},\nabla,\vect{r}_\tau)=&-\frac{1}{\gamma}\nabla F(\vect{r},\vect{r}_\tau)+D\Delta
\end{align}
where $\Delta$ denotes the Laplace operator. The Smoluchowski equation then reads
\begin{align}
&\frac{\partial w(\vect{r},t | \vect{r}_\tau,\tp)}{\partial t} \Bigg|_{\tp=t-\tau}\nonumber\\&\qquad=\frac{1}{\gamma}\nabla\left[k_BT\nabla-\vect{F}(\vect{r},\vect{r}_\tau)\right]w(\vect{r},t | \vect{r}_\tau,\tp)\,.\label{eq_Smolu2_1p}
\end{align}

From this equation, the joint probability $w(\vect{r},t ; \vect{r}_\tau,\tp)=w(\vect{r},t | \vect{r}_\tau,\tp)w(\vect{r}_\tau,\tp)$ is obtained by multiplication with $w(\vect{r}_\tau,\tp)$ as
\begin{align}
&\frac{\partial w(\vect{r},t ;\vect{r}_\tau,\tp)}{\partial t}\Bigg|_{\tp=t-\tau}\nonumber\\&\qquad=\frac{1}{\gamma}\nabla\left[k_BT\nabla-\vect{F}(\vect{r},\vect{r}_\tau)\right] w(\vect{r},t ;\vect{r}_\tau,\tp)\,.
\end{align}
Finally, integration over the past position $\vect{r}_\tau$ yields an expression for the probability \mbox{$w(\vect{r},t)=\int \dif\vect{r}_\tau w(\vect{r},t ;\vect{r}_\tau,\tp)$} to find a particle at $\vect{r}$ at time $t$ without specifying its past position \cite{Frank2005,Frank2005_PRE72,Loos2017,Loos2019,Guillouzic1999,Atay2010}; i.e.,~it leads from the joint probability to have a particle at position $\vect{r}$ at time $t$ and at position $\vect{r}_{\tau}$ at time $\tp$ to one for the joint probability to have a particle at position $\vect{r}$ at time $t$ and at any position at time $\tp$. The resulting probability $w(\vect{r},t)$ is then no longer dependent on the past time $\tp$ and its time evolution is given by
\begin{align}
\frac{\partial w(\vect{r},t)}{\partial t}=&D\Delta w(\vect{r},t )\nonumber\\&
-\frac{1}{\gamma}\int \dif\vect{r}_\tau\,\nabla \vect{F}(\vect{r},\vect{r}_\tau)w(\vect{r},t ;\vect{r}_\tau,\tp)\,.\label{eq_Frank_w}
\end{align}

\subsection{Extension to the many-particle case}
We now return to the original problem of finding a Smoluchowski equation equivalent to Eq.~(\ref{eq_eom_manyp}). In this case, the system state $X(t)$ is given by a set of particle positions which should still obey Eq.~(\ref{eq_Smolu_Frank}). The many-body conditional probability density $w(\vect{r}^N,t | \vect{r}^N_\tau,\tp)$ to have $N$ particles at positions $\vect{r}^N=\vect{r}_1,\cdots,\vect{r}_N$ at time $t$ under the condition that the previous corresponding positions were $\vect{r}^N_\tau=\vect{r}_{\tau_1},\cdots,\vect{r}_{\tau_N}$ at time $\tp=t-\tau$ is thus given by
\begin{align}
&\frac{\partial w(\vect{r}^N,t | \vect{r}_\tau^N,\tp)}{\partial t}\Bigg|_{\tp=t-\tau}\nonumber\\ &=\frac{1}{\gamma}\sum_{i=1}^{N}\nabla_i\left[k_BT\nabla_i-\left(-\nabla_i U_\mathrm{tot}(\vect{r}^N,\vect{r}^N_\tau)\right)\right]w(\vect{r}^N,t | \vect{r}^N_\tau,\tp)\,.\label{eq_Smolu2}
\end{align}
Here, the gradient or divergence $\nabla_i$ indicates differentiation with respect to $\vect{r}_i$. The forces on the particles are expressed as potential gradients $-\nabla_i U_\mathrm{tot}(\vect{r}^N,\vect{r}^N_\tau)$ with the derivative in this term only intended to act on $U_\mathrm{tot}$ and not on $w(\vect{r}^N,t | \vect{r}^N_\tau,\tp)$. The potential includes the direct pair interaction potential between the particles, $\VYu$, and a contribution due to the feedback potential, $\Vext$:
\begin{align}
U_\mathrm{tot}\left(\vect{r}^N,\vect{r}_\tau^{N}\right)=&\frac{1}{2}\sum_{\substack{i,j=1\\i\ne j
}}^{N}\VYu(|\vect{r}_i-\vect{r}_j|)\nonumber\\&+\sum_{i,j=1}^{N}\Vext(|\vect{r}_i-\vect{r}_{\tau_j}|)\,.
\end{align}

Analogously to the one-particle case, multiplication of Eq.~(\ref{eq_Smolu2}) with the probability $w(\vect{r}_\tau^N,\tp)$ to have had $N$ particles at the positions $\vect{r}_\tau^N$ at time $\tp$ leads to an equation for the joint probability $w(\vect{r}^N,t ;\vect{r}_\tau^N,\tp)=w(\vect{r}^N,t | \vect{r}_\tau^N,\tp)w(\vect{r}_\tau^N,\tp)$, given by
\begin{align}
&\frac{\partial w(\vect{r}^N,t ;\vect{r}_\tau^N,\tp)}{\partial t}\Bigg|_{\tp=t-\tau}\nonumber \\
&=\frac{1}{\gamma}\sum_{i=1}^{N}\nabla_i\left[k_BT\nabla_i+\left(\nabla_i U_\mathrm{tot}(\vect{r}^N,\vect{r}_\tau^N)\right)\right] w(\vect{r}^N,t ;\vect{r}_\tau^N,\tp)\,.\label{eq_Smolu3}
\end{align}
Integration over the past positions $\vect{r}_\tau^N$ yields an equation for the probability density solely dependent on the set of current positions $\vect{r}^N$ given by
\begin{align}
w(\vect{r}^N,t)=\int \dif\vect{r}_{\tau_1}\cdots\int \dif\vect{r}_{\tau_N} w(\vect{r}^N,t ;\vect{r}_\tau^N,\tp)\,.
\end{align} 
Performing the integration, the potential term in Eq.~(\ref{eq_Smolu3}), 
\begin{align}
&\int \dif\vect{r}_{\tau_1}\cdots\int \dif\vect{r}_{\tau_N} U_\mathrm{tot}(\vect{r}^N,\vect{r}_\tau^N)  w(\vect{r}^N,t ;\vect{r}_\tau^N,\tp)\nonumber\\&= \int \dif\vect{r}_{\tau_1}\cdots\int \dif\vect{r}_{\tau_N} U_\mathrm{tot}(\vect{r}^N,\vect{r}_\tau^N) w(\vect{r}^N,t | \vect{r}_\tau^N,\tp) w(\vect{r}_\tau^N,\tp)\,,
\end{align}
cannot be traced back to a simple expression due to the fact that both $U_\mathrm{tot}(\vect{r}^N,\vect{r}_\tau^N)$ and 
$w(\vect{r}^N,t | \vect{r}_\tau^N,\tp)$ depend on the previous particle positions $\vect{r}_\tau^N$. 
Intuitively, this can be understood as being due to the coupling of $U_\mathrm{tot}(\vect{r}^N,\vect{r}_\tau^N)$ to the particle trajectories. This renders the reduction of the potential to a single value at the current position impossible.
The many-particle equivalent of Eq.~(\ref{eq_Frank_w}) obtained from integration of Eq.~(\ref{eq_Smolu3}) thus reads
\begin{align}
\frac{\partial w(\vect{r}^N,t)}{\partial t}=& D\sum_{i=1}^{N}\Delta_i w(\vect{r}^N,t )
+\frac{1}{\gamma}\int \dif\vect{r}_{\tau_1}\cdots\int \dif\vect{r}_{\tau_N}\nonumber\\  &\times\sum_{i=1}^{N}\nabla_i w(\vect{r}^N,t ;\vect{r}_\tau^N,\tp)\nabla_i U_\mathrm{tot}(\vect{r}^N,\vect{r}_\tau^N)\label{eq_Smolu_w}\,.
\end{align}

Next, we define the one- and two-particle densities, $\rho$ and $\rho^{(2)}$, as well as the time-shifted two-particle density, $\rho^{(2)}_\mathrm{s}$, via
{\allowdisplaybreaks
\begin{align}
&\rho\left(\vect{r}_1,t\right)=N\int \dif\vect{r}_2\cdots\int \dif\vect{r}_N\nonumber\\&\qquad\qquad\int \dif\vect{r}_{\tau_1}\cdots\int \dif\vect{r}_{\tau_N}\ w\left(\vect{r}^N,t;\vect{r}_{\tau}^{N},\tp\right)\\
&\rho^{(2)}\left(\vect{r}_1,\vect{r}_2,t\right)=N\left(N-1\right)\int \dif\vect{r}_3\cdots\int \dif\vect{r}_N\nonumber\\ &\qquad\qquad\int \dif\vect{r}_{\tau_1}\cdots\int \dif\vect{r}_{\tau_N}\ w\left(\vect{r}^N,t; \vect{r}_{\tau}^{N},\tp\right)\\
&\rho^{(2)}_\mathrm{s}(\vect{r}_1,t; \vect{r}_{\tau},\tp)=N\int \dif\vect{r}_2\cdots\int \dif\vect{r}_N\nonumber\\ 
&\qquad\int \dif\vect{r}_{\tau_1}\cdots\int \dif\vect{r}_{\tau_N}\  w\left(\vect{r}^N,t;\vect{r}_{\tau}^{N},\tp\right)\delta(\vect{r}_\tau-\vect{r}_{\tau_1})\nonumber\\ 
&\quad+N(N-1)\int \dif\vect{r}_2\cdots\int \dif\vect{r}_N\nonumber\\
&\quad\ \ \int \dif\vect{r}_{\tau_1}\cdots\int \dif\vect{r}_{\tau_N}\ 
 w\left(\vect{r}^N,t;\vect{r}_{\tau}^{N},\tp\right)\delta(\vect{r}_\tau-\vect{r}_{\tau_2})\,.
\end{align}}
Here, the instantaneous densities $\rho\left(\vect{r}_1,t\right)$ and $\rho^{(2)}\left(\vect{r}_1,\vect{r}_2,t\right)$ do not depend on $\tp$ due to integration over the past positions $\vect{r}_{\tau}^N$. 
Furthermore, note that $\rho^{(2)}_\mathrm{s}(\vect{r}_1,t;\vect{r}_{\tau},\tp)$ includes a self-term (first part) as well as a term due to other particles (second part) such that the resultant two-particle density is defined as the probability to find one particle at $\vect{r}_1$ at time $t$ and any particle (i.e.,~the same or another) at position $\vect{r}_{\tau}$ at time $t_\tau$.

To obtain an equation for the evolution of the one-particle density, Eq.~(\ref{eq_Smolu_w}) is integrated with $N\int \dif\vect{r}_2\cdots\int \dif\vect{r}_N$. Using the above definitions, this leads to the exact equation
\begin{align}
\gamma &\frac{\partial\rho\left(\vect{r}_1,t\right)}{\partial t}
= k_BT\Delta_1\rho\left(\vect{r}_1,t\right)\nonumber\\
&+\int \dif\vect{r}_2\nabla_1\rho^{(2)}\left(\vect{r}_1,\vect{r}_2,t\right)\nabla_1\VYu(|\vect{r}_1-\vect{r}_2|)\nonumber\\
&+\int \dif\vect{r}_{\tau}\nabla_1\rho_\mathrm{s}^{(2)}\left(\vect{r}_1,t;\vect{r}_{\tau},\tp\right)\nabla_1\Vext(|\vect{r}_1-\vect{r}_{\tau}|)\,.\label{eq_dens1}
\end{align}
It is worth mentioning that the last term, i.e.,~the interaction via the feedback potential, is similar to the preceding one, meaning that it takes the same form as a pair potential. The two terms differ in that the feedback does not act between positions at the same time but positions at time $t$ and previous ones at time $\tp$. Also, the last term includes a self-term which is not present in the direct interactions.

\subsection{Derivation of the DDFT equation}
Starting from the many-body time-delayed Smoluchowski equation [Eq.~(\ref{eq_dens1})], which is 
stochastically equivalent to the Langevin equation [Eq.~(\ref{eq_eom_manyp})], we construct a dynamical density functional theory (DDFT) \cite{Archer2004}. The DDFT is then used in the next section to investigate the stability of the homogeneous density state against wave formation.

Equation~(\ref{eq_dens1}) is exact but not self-contained as it includes the instantaneous and the time-shifted two-particle density. To obtain a closed equation, we first rewrite the direct interaction potential using the adiabatic approximation of DDFT,
\begin{equation}
\int \dif\vecr_2 \rho^{(2)}\left(\vecr_1, \vecr_2\right)\nabla_1 \VYu(|\vect{r}_1-\vect{r}_2|)=-k_BT\rho\left(\vecr_1\right)\nabla_1 c^{(1)}\left(\vecr_1\right)\,,
\end{equation}
which expresses the particle interaction forces via the direct correlation function $c^{(1)}(\vecr)$. The latter is related to the excess free energy, i.e., the nonideal part of the free energy, $\mathcal{F}_\text{exc}$, by
\begin{equation}
c^{(1)}\left(\vecr\right)=-\frac{1}{k_BT}\frac{\delta \mathcal{F}_\text{exc}\left[\rho(\vect{r})\right]}{\delta\rho\left(\vect{r}\right)}\label{eq_c1}
\end{equation}
in equilibrium. The adiabatic approximation assumes the same relation in nonequilibrium.

Second, it is necessary to focus on the time-shifted two-particle density.
In fact, even for the noninteracting particle case ($\VYu=0$) this term leads to a hierarchy of equations whose solution is not at all trivial and subject to current research \cite{Frank2005,Frank2005_PRE72,Loos2017,Loos2019,Guillouzic1999,Atay2010}. Here, we use the mean-field-like approximation
\begin{align}
\rho_\mathrm{s}^{(2)}\left(\vect{r}_1,t;\vect{r}_{\tau},\tp\right)&\approx\rho\left(\vect{r}_1,t\right)\rho\left(\vect{r}_{\tau},\tp\right)\nonumber\\&\quad=\rho\left(\vect{r}_1,t\right)\rho\left(\vect{r}_{\tau},t-\tau\right)\,.
\end{align}
At this point, we are thus neglecting the correlations between the present and past positions of the particles. The description is then equivalent to having particles move in an effective external potential
\begin{equation}
\bar{V}_\mathrm{fb}(\vect{r}^N,t)=\int \dif\vect{r}_\tau\rho\left(\vect{r}_\tau,t-\tau\right)\Vext(|\vect{r}-\vect{r}_\tau|)\label{eq_effPot}\,,
\end{equation} 
which is obtained through a convolution of the time-shifted one-particle density with the feedback potential $\Vext$.
This is a crude approximation for systems where the different sets of actual particle paths are important and the one-particle density is not sufficient to describe the system.
It should, however, be decent if we are close to a homogeneous state due to a small amplitude of the feedback potential or if we consider only a small perturbation to this state, as done in the next section. Such a homogeneous reference state is expected to lose its memory of the specific realization of past positions on the time scale $1/D\rho_0$, i.e.,~the time needed for particles to diffuse the mean interparticle distance. 
However, this approximation is in particular \textit{not} well justified for dilute feedback systems or very short delay times $\tau\ll 1/D\rho_0$.

Finally, inserting the above approximations into Eq.~(\ref{eq_dens1}), a self-contained approximative equation for the one-particle density is obtained:
\begin{align}
\gamma&\frac{\partial\rho\left(\vect{r},t\right)}{\partial t}=k_BT\Delta\rho\left(\vect{r},t\right)-k_BT\nabla\rho(\vect{r})\nabla c^{(1)}(\vect{r})\nonumber\\
&+\nabla\rho\left(\vect{r},t\right)\nabla\int \dif\vecr_\tau\rho\left(\vecr_\tau,t-\tau\right)\Vext(|\vect{r}-\vecr_\tau|)\,.\label{eq_DDFT}
\end{align}

\subsection{Linear stability analysis}
Next, the effect of a small perturbation $\rhop(\vect{r},t)$ on a homogeneous state of constant density $\rho_0$ is investigated with
\begin{equation}
\rho(\vect{r},t)=\rho_0+\rhop(\vect{r},t)
\end{equation} 
being the space- and time-dependent density. We expand the direct correlation function $c^{(1)}(\vecr)$ about the bulk fluid value $\rho_0$ up to linear order,
\begin{align}
c^{(1)}(\vecr)&\approx c^{(1)}_0+\int \dif\vecr_2 \frac{\delta c^{(1)}(\vecr)}{\delta \rho\left(\vecr_2\right)}\bigg\vert_{\rho_0}\rhop(\vect{r}_2,t)\nonumber\\&= c^{(1)}_0+\int \dif\vecr_2\ c^{(2)}\left(|\vecr-\vecr_2|;\rho_0\right)\rhop(\vect{r}_2,t)\,,
\end{align}
with the direct pair-correlation function
\begin{equation}
c^{(2)}\left(|\vecr-\vecr_2|;\rho_0\right)=-\frac{1}{k_BT}\frac{\delta^2 \mathcal{F}_\text{exc}\left[\rho(\vect{r})\right]}{\delta\rho\left(\vect{r}_2\right)\delta\rho\left(\vect{r}\right)}\bigg\vert_{\rho_0}
\end{equation}
and insert this expression into Eq.~(\ref{eq_DDFT}). Linearization of the result in the perturbation density $\rhop(\vect{r},t)$ and nondimensionalization yields the equation 
\begin{eqnarray}
\frac{\partial\tilde{\rho}\left(\vect{k},t\right)}{\partial t}=-k^2\Big[ \tilde{\rho}\left(\vect{k},t\right)-\rho_0\,\tilde{\rho}(\vect{k},t)\tilde{c}(k;\rho_0)\nonumber\\
+\rho_0\,\tilde{\rho}\left(\vect{k},t-\tau\right)\Vextt(k)\Big]\label{eq_DDFT_pert_FT_u}\,,
\end{eqnarray}
where $\tilde{\rho}\left(\vect{k},t\right)$ indicates the Fourier transform of $\rhop(\vect{r},t)$,  $\tilde{c}(k;\rho_0)$ the one of $c^{(2)}\left(|\vecr-\vecr_2|;\rho_0\right)$, and $\Vextt$ the one of the feedback potential.

Taking the ansatz for a wave solution of wave vector $\vect{k}$ and amplitude $\varepsilon$,
\begin{equation}
\rhop(\vect{r},t)=\varepsilon\ \rme^{i\vect{k}\vect{r}}\rme^{\lambda t}\,,
\end{equation} and inserting it into Eq.~(\ref{eq_DDFT_pert_FT_u}), we obtain the dispersion relation
\begin{equation}
\lambda=-k^2\Big[ 1-\rho_0\,\tilde{c}(k;\rho_0)+\rho_0\,\rme^{-\lambda\tau}\Vextt(k)\Big]\label{eq_DDFT_pert_FT_k}\,.
\end{equation}
The solution for $\lambda$ can be separated into a real part $\alpha$ describing the growth of the perturbation and an imaginary part $\omega$ giving the angular frequency of the traveling wave as 
\begin{equation}
\lambda(k)=\alpha(k)+i\omega(k)\,.
\end{equation} 
This gives the two equations
\begin{align}
\alpha(k)&=-k^2\Big[ 1-\rho_0\,\tilde{c}(k;\rho_0)+\rho_0\,\rme^{-\alpha(k)\tau}\cos(\omega(k)\tau)\Vextt(k)\Big]\,,\nonumber\\
\omega(k)&=k^2\rho_0\,\rme^{-\alpha(k)\tau}\sin(\omega(k)\tau)\Vextt(k)\,,\label{eq_DDFT_solution}
\end{align}
which implicitly define the solution. Since $\tilde{c}(k;\rho_0)$ for the equilibrium bulk fluid is required here as an input, we use a reference simulation without the feedback potential to obtain $\tilde{c}(k;\rho_0)$ which is related to the structure factor of the system by \cite{book_Hansen1990}
\begin{equation}
\tilde{c}(k;\rho_0)=\rho_0\left(1-\frac{1}{S(k)}\right)\,.
\end{equation}

\FloatBarrier

\subsection{DDFT results and comparison to simulations}
\begin{figure}[tbh]
\includegraphics[width=0.99\columnwidth]{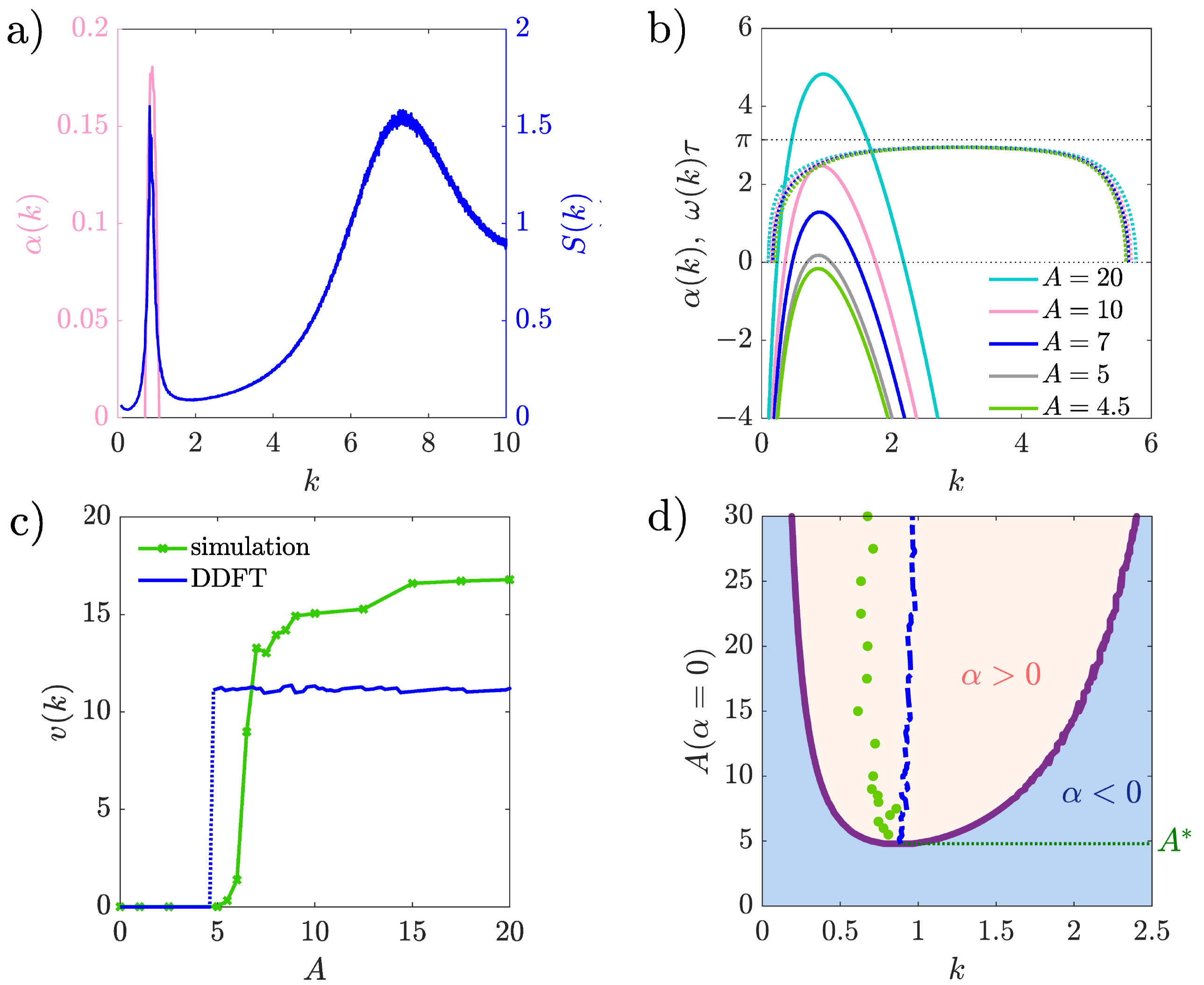}
\caption{Prediction of the instability using DDFT: (a) Simulation result for structure factor $S(k)$ and DDFT prediction for the growth rate $\alpha(k)$ for $A=5$. (b) Growth rate $\alpha(k)$ and angular frequency $\omega(k)$ for different strengths of the feedback potential. Solid lines give values for $\alpha$, dotted ones those for $\omega\tau$. (c) Comparison between DDFT prediction of band velocity and simulation results. (d) Stability curve giving the wave-number-dependent value of $A$ for which the growth factor $\alpha$ changes from negative to positive. The minimal value of $A$, for which an $\alpha=0$ exists, is denoted by $A^*\approx4.8$. The green dots show the wave number at which patterning was observed in the simulations. The blue dashed line shows the maximally unstable wavelength according to the DDFT prediction.\label{fig_DDFT}}
\end{figure}

Results for $\alpha(k)$ and $\omega(k)$ are shown in Fig.~\ref{fig_DDFT}. Figure~\ref{fig_DDFT}(a)
displays the range of unstable wave numbers defined via a positive $\alpha (k)$ and puts these into relation to the structure factor $S(k)$.
Indeed, the maximally unstable wave number, i.e., the wave number at which $\alpha (k)$ has its maximum, coincides with the structural bandwidth characterized by $k^*$.

Figure~\ref{fig_DDFT}(b) shows the full dispersion $\alpha (k)$ and $\omega(k)$ for various $A$.
The range of unstable wave numbers is zero below the transition ($A<4.8$) and increases 
with the feedback amplitude $A$ above it. 
From the frequency $\omega(k)$, close to the transition, the phase velocity of the imposed wave is determined by $\omega(k)/k$. At $k=k^*$ this phase velocity should be close to the band velocity $v$ such that the band velocity predicted by the theory is
\begin{equation}
v = \frac{\omega(k^*)}{k^*}\,.\label{eq_vomegak}
\end{equation}
According to Eqs.~(\ref{eq_vtheo}) and (\ref{eq_vomegak})
we obtain an approximative phase shift of $\omega(k^*)\tau = \pi$ within the delay time. In fact, Fig.~\ref{fig_DDFT}(b) reveals that the
phase shift $\omega(k^*)\tau$ is close to $\pi$ but deviations appear due to the approximative nature of Eq.~(\ref{eq_vtheo}). The phase shift $\omega(k^*)\tau$
is close to the ideal value $\pi$ for any $k$ close to the maximally unstable wave vector with discrepancies indicating that the wave does not change exactly between a pattern and its negative image after a delay time $\tau$.

The theoretical prediction of the band velocity is compared to the simulation data in Fig.~\ref{fig_DDFT}(c) and reasonable agreement is found with respect to onset and magnitude of the velocity as a function of the feedback potential amplitude $A$.

Figure~\ref{fig_DDFT}(d) illustrates the stability in the plane defined by the wave number $k$ and the feedback amplitude $A$. 
There is a separatrix distinguishing a stable [$\alpha (k)<0$]
from an unstable [$\alpha (k)>0$] regime. The instability occurs beyond a critical amplitude $A^*\approx 4.8$ with an instability wave number $k^*=0.87$.
The DDFT prediction for the critical amplitude $A^*\approx 4.8$ compares favorably with the simulation results of $5 \le A^*\le 6$
which documents the predictive power of the microscopic theory. 
Deviations may be due to
the approximative pair correlation function entering as an input for the nonequilibrium steady state or due to the adiabatic approximation. 

In the unstable regime, $\alpha (k)>0$, the maximally unstable wave number, i.e.,~the one with the fastest growth rate, is shown by the blue dashed line in Fig.~\ref{fig_DDFT}(d). This prediction can be compared to the actual band wave number $k^*$ found in the simulations
shown as green dots. While at the onset of the transition these two wave numbers are very close [see also Fig.~\ref{fig_DDFT}(a)], they exhibit a different trend for increasing $A$ indicating that nonlinear effects play a larger role away from criticality.

\FloatBarrier
 
\section{Circular confinement}\label{sec_confine}
In the case of periodic boundary conditions as discussed in Sec.~\ref{sec_simu} the commensurability of the box size and the (projected) wavelength effectively constrains the possible band orientations. Inversely, this constraint may affect the actually realized wavelength as certain combinations of orientation and wavelength that fit well with the box size may be preferred. In the following, we aim to investigate this effect in more detail by confining the system to a prescribed geometry.
We choose a circular confinement which is imposed
onto the system by an external potential $\VYu_\mathrm{c}(r)$ which is radially symmetric. The potential is soft but diverges at a 
distance $r=R$ from the center such that a circular confinement is realized. We choose the external potential form
\begin{equation}
\VYu_\mathrm{c}(r)=\frac{V_0}{R-r}\ \rme^{-\kappa (R-r)}
\end{equation}
and fix the radius to $R\approx45\,b$ such that the density $\rho_0=N/\pi R^2=1/b^2$ with $N=6400$ is the same as in the previous case.

\begin{figure}[h]
\includegraphics[width=0.99\columnwidth]{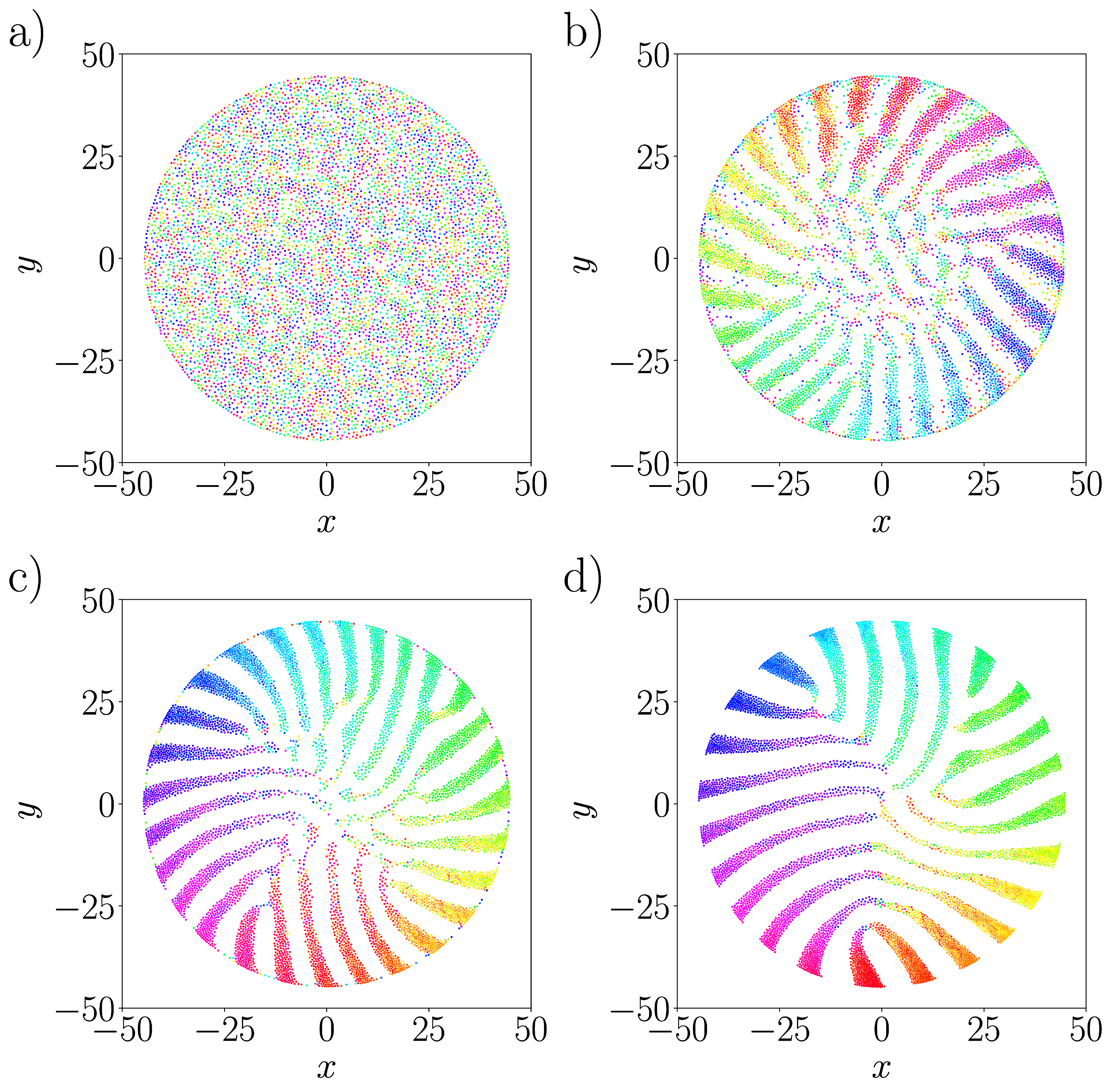}
\flushright
\includegraphics[width=0.6\columnwidth]{colorbar_phi.pdf}
\caption{Particle positions for potential strength (a) $A=5.5$, (b) $A=7$, (c) $A=10$, and (d) $A=20$ at time $t=500$. The color code is the same as used in Fig.~\ref{fig_snapshot_t500} and indicates the direction of movement of the particles. \label{fig_circularconfined}}
\end{figure}

Representative snapshots of the system for different feedback potential amplitudes $A$ are shown in Fig.~\ref{fig_circularconfined}.
Band formation is found as for the nonconfined case, with the pattern reaching further inwards if the feedback potential strength is increased. This is because the confining boundaries set a preferred band orientation and thus facilitate the formation of a traveling wave along the boundary.
Due to the circular form of the confining potential, the bands are now forced into a rotational motion. Moreover, the circular geometry does not allow parallel wavefronts with a constant separation running from the boundary to the center. Thus, defects are unavoidable.
At low and intermediate feedback potentials the formed pattern is approximately rotationally symmetric with respect to the center but not very well ordered. In contrast, at high values of $A$, the waves are well-established and have an about constant wavelength. However, to allow for this arrangement, a separation into domains of different band orientation with a complicated topology at the center is required and observed.

\begin{figure}[htb]
\includegraphics[width=0.95\columnwidth]{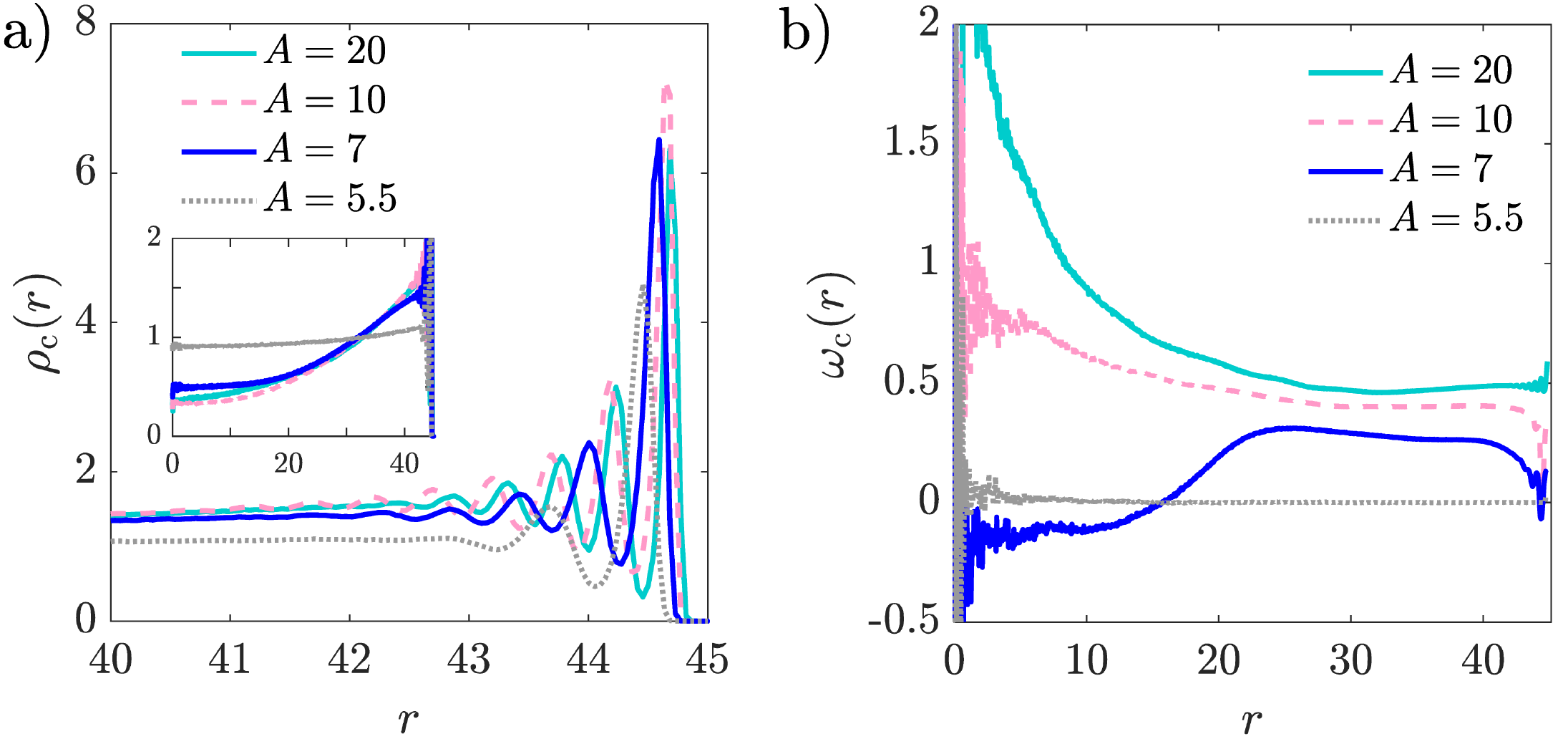}
\caption{(a) Radial density profile $\rho_\mathrm{c}(r)$ and (b) angular drift velocity profile $\omega_\mathrm{c}(r)$. 
In (b) the sign has been adjusted such that $\omega_\mathrm{c}(r)$ takes a positive value for the particles in the bands (i.e.,~in the range $r\approx20-40$).
The feedback potential leads to a density accumulation at the system boundary and a rotational motion around the center.
\label{fig_circulardens}}
\end{figure}

To further examine the system response in this case, we consider the time-averaged radial density profile 
\begin{equation}
\rho_\mathrm{c}\left(r\right)= \Big\langle \sum_{i=1}^N \delta\left(\vect{r}-\vect{r}_i(t') \right)\Big\rangle\,,
\end{equation}
which by symmetry depends only on the radial distance $r$ to the center.
Next, we  also define a time-averaged angular drift velocity profile $\omega_\mathrm{c}(r)$:
\begin{align}
\omega_\mathrm{c}(r)=\hat{\mathbf{e}}_z\cdot\Big\langle\sum_{i=1}^{N}\frac{\vect{r}_i(t')\times\vect{v}_i(t')}{\vect{r}_i^2(t')}\delta\left(\vect{r}-\vect{r}_i(t')\right)\Big\rangle\frac{1}{\rho_\mathrm{c}(r)}
\end{align}
where $\hat{\mathbf{e}}_z$ is the unit vector perpendicular to the plane of motion. The profile $\omega_\mathrm{c}(r)$ likewise depends on 
the radial distance to the center only.
Figure~\ref{fig_circulardens} shows results for the density profile $\rho_\mathrm{c}(r)$ and the angular velocity $\omega_\mathrm{c}(r)$. 
For growing feedback potential strength $A$, an accumulation of particles near the system boundary is observed 
as signaled by a peak in the density profile close to the wall accompanied by a (relative) depletion of particles near the center.
This effect has also been found for self-propelled particles in circular confinement \cite{Smallenburg_Loewen_PRE_2015,Naji2018}
and can qualitatively be understood here on similar grounds as suggested by the MSD (Fig.~\ref{fig_MSD}) which resembles active Brownian particles.
The difference between the two cases is that here, the individual particle velocity is typically directed tangentially to the wall. Therefore, in the feedback case the accumulation is driven by wall curvature and larger $A$ enhances the wall accumulation effect. In contrast, in the active particle case it is driven by persistence in the otherwise unconstrained propulsion direction.

Considering the angular velocity, we find that for small feedback potentials it is almost zero at all distances $r$ from the center. 
Increasing $A$ leads to the formation of a two-layer state with the inner part revolving in the \textit{opposite} direction to the outer one, 
signaled by a sign change in $\omega_\mathrm{c}$. The continuous change in $\omega_\mathrm{c}$ with distance $r$ is consistent with a motion spiraling outwards. 
At even higher values of $A$, this feature disappears and all particles rotate like a rigid body with a constant joint angular velocity apart from some remaining distance-dependence in the central part of the profile. Thus, the slope of $\omega_\mathrm{c}$
is negative, implying a motion which is spiraling inwards.
Furthermore, close to the wall, particles move in the opposite 
direction to the bands or at least significantly more slowly than the band. This is especially observed for intermediate feedback strengths, 
signaled by a dip in the angular velocity to values 
below zero. This effect is interpreted as an escape of particles from the
ideal bands induced by wall curvature, leading to counterpropagating particles which are visible in the system snapshots of Figs.~\ref{fig_circularconfined}(b) and (c). It vanishes again for strong feedback strengths [see Fig.~\ref{fig_circularconfined}(d)]. 

\begin{figure}[tbh]
\includegraphics[width=0.99\columnwidth]{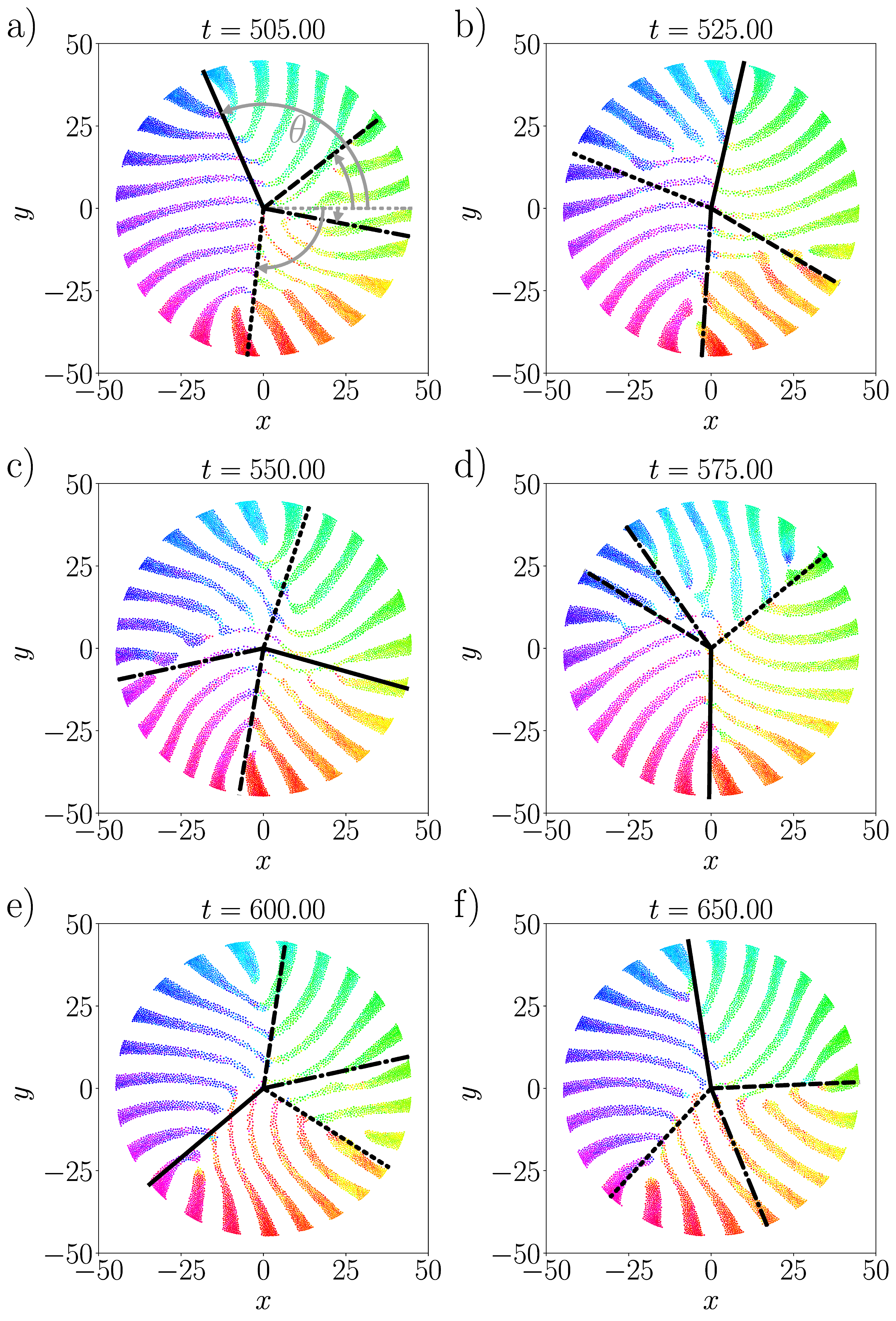}
\flushright
\includegraphics[width=0.6\columnwidth]{colorbar_phi.pdf}
\caption{Snapshots for potential amplitude $A=20$ 
for approximately one full rotation of the slow domain boundaries which occurs over a time of about 150. 
As in Fig.~\ref{fig_snapshot_t500}, colors indicate the direction of the particle drift velocity. 
The phase boundaries between the different domains are also indicated as black lines. Their movement with respect to time is measured by the angles $\theta$ [indicated in (a)] and plotted in Fig.~\ref{fig_angles_phases}.}
\label{fig_snapshots_phases}
\end{figure}

For high feedback potential amplitude, the emergence of domains with different band orientations is observed. These domains are separated by 
grain boundaries which, interestingly, are neither static nor co-rotating with the same mean angular speed as the particles. Instead, they rotate with a considerably slower angular speed albeit in the same direction. This is documented in Fig.~\ref{fig_snapshots_phases}, where typical system snapshots are shown
during one full but slow rotation of the system grain boundaries. Four boundaries separating different domains are indicated in Fig.~\ref{fig_snapshots_phases}.

\begin{figure}[tbh]
\includegraphics[width=0.7\columnwidth]{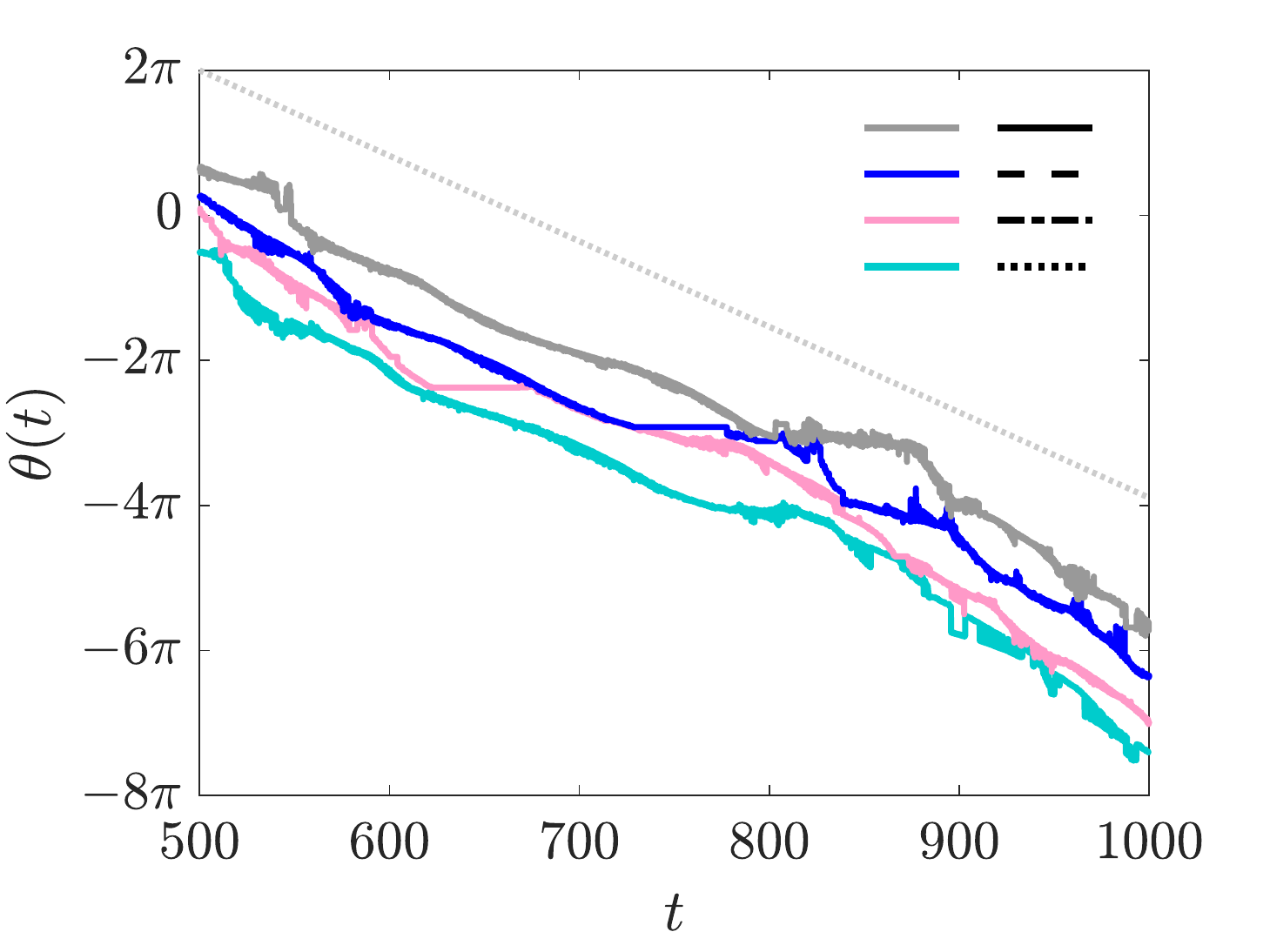}
\caption{Time dependence of the angular position $\theta$ of the grain boundaries indicated by the black lines in Fig.~\ref{fig_snapshots_phases}. For better visibility, the positions of the different grain boundaries are distinguished by color. The corresponding line styles of the boundaries indicated in Fig.~\ref{fig_snapshots_phases} are shown in the legend.
The mean angular velocity of the grains is indicated by the dotted line.
} \label{fig_angles_phases}
\end{figure}

The angular position of the grain boundaries as a function of time is shown in Fig.~\ref{fig_angles_phases}.
Their angular positions change approximately linearly in time.
A linear fit yields an average rotation speed of the grains of $\omega_\mathrm{g}\approx-0.04$ for the 
system of Fig.~\ref{fig_circularconfined}(d), i.e., for a feedback amplitude of $A=20$. This is one order of magnitude smaller than the mean particle angular velocity 
which is  $\omega_\mathrm{c}\approx-0.5$
in the outermost part of the cavity [see Fig.~\ref{fig_circulardens}(b)].
We speculate that the slower grain boundary speed is related to the group velocity
of propagating bands in the bulk which is also much smaller than the phase velocity. In fact, from Fig.~\ref{fig_DDFT}(b), the group velocity $d \omega (k) / dk|_{k=k^*}$
is only 13\% of the phase velocity $\omega(k^*)/k^*$ at the wave number $k^*$ for the given potential strength ($A=20$).

To summarize, the feedback-driven suspension forms moving bands also under circular confinement. Due to the impenetrability of the walls, the band velocity has to be tangential to the wall; i.e., the band lamellae must be oriented perpendicular to the wall if the bands move along the normal of their interface as in the unconstrained case. While confinement to a straight (i.e.,~uncurved) slit geometry supports band formation perpendicular to the walls by imposing a preferred band orientation (data not shown), wall curvature destabilizes such bands due to the incompatibility with straight bands.
Since the constraint of perpendicularity cannot be sustained for a \textit{straight} band along a \textit{curved} wall, the bands must either bend or change their width as a function of radial distance to the center such that $\lambda_\mathrm{b}\equiv\lambda_\mathrm{b}(r)$. Furthermore, as in the periodic boundary case, the fixed length of the circumference also poses a commensurability constraint for the band wavelength. However, this second constraint is weakened due to the grain boundaries. For the large feedback strengths considered here, the system exhibits tilted band formation. Dynamically this implies that 
spiraling waves are formed for which both inward and outward orientation are observed. In particular, particles near the curved wall 
cannot accommodate their individual motion with the global band motion and depending on $A$ are either left behind or propagate faster [see Fig.~\ref{fig_circulardens}(b)]. As a further important consequence,
the convex curvature of the boundary leads to a significant particle accumulation near the wall and a depletion from the system center. The latter constitutes a natural source of disorder as topologically the cavity can never fulfill the constraints
of curved bands meeting in the center. Instead the center becomes a nucleation point for grains with different band orientations. For strong feedback strengths, these grains co-rotate with much smaller angular velocities than that of the individual bands.

\FloatBarrier
\section{Conclusions} \label{sec_conclusion}
We have shown that subjecting colloidal particles to a repulsive feedback potential dependent on their previous positions can lead to pattern  
formation. For a sufficiently strong repulsive potential, we have found a transition to a traveling band phase which can be predicted by a linear stability 
analysis of dynamical density functional theory. Under circular confinement, the transition persists 
but becomes more complex, exhibiting wall accumulation of particles, spiraling patterns, and creation of banded domains. Our findings can be verified in experiments
by using colloids in an external light field \cite{Hanes2009,Evers2013,Volpe2015,Bewerunge2016,Baeuerle2018}  or
employing autochemotactic particles \cite{Jin2017,Jin2018}.

Future work can be performed along the following directions: First, it is worth considering an {\it attractive} feedback potential
with a negative amplitude $A<0$. For attractive feedback, particles tend to stay at their previous positions.
Starting from an initial homogeneous system, this leads to the formation of particle clusters accompanied by a slowing down of the dynamics 
and to subsequent coarsening. The emerging structures are expected to be similar 
to those found in phase separation \cite{Hajime_Tanaka_review}. It would thus be interesting to identify and compare the scaling laws for the mean cluster size as a function of time in different regimes for various system parameters.

Second, while our scenario will not qualitatively change for other types of soft repulsion between the particles, a long-ranged attraction \cite{OlsonReichhard2010,Zhao2012} in the interparticle forces can give rise to equilibrium
gas-liquid phase separation \cite{Vliegenhard} with a critical point. If a repulsive feedback potential is applied,
there is competition between bulk phase separation
and band formation which may lead to interesting new structures.

Third, if one considers higher particle densities or lower temperature than in this paper, the equilibrium two-dimensional Yukawa system exhibits freezing into a hexatic or hexagonal crystalline phase \cite{Kapfer_Krauth}. Applying a repulsive feedback 
is then expected to lead to the formation of traveling crystalline bands (``solids''), the detailed structure of which still has to be worked out.
The density increase in the bands induced by the feedback will additionally support and enhance traveling crystal formation. Thus, it is expected that the phase boundary will depend on the amplitude $A$.
Dynamical density functional theory can in principle be applied to describe crystallization in nonequilibrium \cite{van_Teeffelen,Zimmermann}.

Fourth, active particles in feedback potentials present a rich playground to investigate further complex collective effects due to competition between activity and feedback forces. Even in the absence of feedback
the collective behavior of active particles is rich \cite{Krauth} and it remains to be explored how swarming \cite{Winkler} and motility-induced phase separation \cite{Tailleur} compete with feedback potentials.
Again dynamical density functional theory can be employed for an appropriate instability analysis \cite{Menzel_Saha_JCP_2016}. 

Last, possible future work should also include extending the system presented here to three dimensions and considering the effect of more 
complex confining geometries \cite{Beppu2017}, for example, nonconvex walls and moving boundaries. Likewise, multiple (competing) \cite{Ahlborn2005} or more complex \cite{Ciszak2015,Pyragas2018} feedback terms can be considered. The latter may also take a density-dependent form, 
thus describing quorum sensing \cite{Baeuerle2018}.

\vspace*{0.5cm}
\begin{acknowledgments}
We thank W. Zimmermann, B. Liebchen, C. Hoell and M. Tarama for helpful discussions.
We gratefully acknowledge financial support by the DFG Grants No.~\mbox{LO 418/19-1} and No.~\mbox{EG 269/6-1}.
\end{acknowledgments}

\FloatBarrier

\bibliography{references_final}

\end{document}